\documentclass{aa}

\usepackage{graphicx}
\usepackage{natbib}
\usepackage{ulem}
      \def\new#1 {#1}
      \def\cut#1 {}

\def\pow#1#2{#1$\times$10$^{#2}$}

\def\hii{H~{\sc II}}

\def\hh{H$_2$}

\def\hho{H$_2$O}
\def\hhoe{H$_2^{18}$O}
\def\hhhop{H$_3$O$^+$}

\def\meth{CH$_3$OH}

\def\cchh{C$_2$H$_2$}

\def\soo{SO$_2$}

\def\hhhp{H$_3^+$}

\def\smm{(sub-)mil\-li\-me\-ter}

\def\farcs{\mbox{$.\!\!^{\prime\prime}$}}

\def\gtsim{{_>\atop{^\sim}}}
\def\ltsim{{_<\atop{^\sim}}}
\def\vlsr{V$_{\rm LSR}$}

\def\dv{$\Delta$\textit{V}}
\def\kms{km~s$^{-1}$}
\def\kkms{K~km~s$^{-1}$}
\def\tmb{$T_{\rm MB}$}
\def\txc{$T_{\rm ex}$}

\def\scm{cm$^{-2}$}
\def\ccm{cm$^{-3}$}

\def\ccms{cm$^{-3}$~s$^{-1}$}
\def\mic{$\mu$m}

\def\msol{M$_{\odot}$}
\def\lsol{L$_{\odot}$}

\def\aap{{A\&A}}

\def\apjl{{ApJ}}
\def\apjs{{ApJS}}

\renewcommand{\citep}[1]{(\citeauthor{#1} \citeyear{#1})}

\begin{document}

\title{\cut{The chemistry of} Water in the
  envelopes and disks \cut{and outflows } around young high-mass
  stars}

\author{F. F. S. van der Tak\inst{1} \and C. M. Walmsley\inst{2} \and F. Herpin\inst{3}
        \and C. Ceccarelli\inst{4}}
\institute{
  Max-Planck-Institut f\"ur Radioastronomie, Auf dem H\"ugel 69, 53121 Bonn, Germany;
  \email{vdtak@mpifr-bonn.mpg.de}
  \and Osservatorio Astrofisico di Arcetri, Largo E.\ Fermi 5, 50125 Firenze, Italy
  \and Observatoire de Bordeaux, L3AB, UMR 5804, B.P.\ 89, 33270 Floirac, France
  \and Laboratoire Astrophysique de l'Observatoire de Grenoble, BP 53, 38041 Grenoble, France}

\titlerunning{Water in regions of high-mass star formation}
\authorrunning{Van der Tak et al.}

\date{Received 28 July 2005 / Accepted 21 October 2005}

\abstract{Single-dish spectra and interferometric maps of \smm\ lines
  of \hhoe\ and HDO are used to study the chemistry of water in eight
  regions of high-mass star formation.
  The spectra indicate HDO excitation temperatures of $\sim$110~K and
  column densities \new{in an 11$''$ beam } of $\sim$2$\times$10$^{14}$~\scm\ for HDO and
  $\sim$2$\times$10$^{17}$~\scm\ for \hho, 
  with the $N$(HDO)/$N$(\hho) ratio increasing with decreasing temperature.
  Simultaneous observations of \meth\ and \soo\ indicate that 20 -- 50\%
  of the single-dish line flux arises in the molecular outflows of
  these objects. \new{The outflow contribution to the \hhoe\ and HDO emission is
  estimated to be 10 -- 20\%. }
  Radiative transfer models indicate that the water abundance is low
  ($\sim$10$^{-6}$) outside a critical radius corresponding to a
  temperature in the protostellar envelope of $\approx$100~K, and `jumps' to
  \hho/\hh$\sim$10$^{-4}$ inside this radius. This value corresponds to the
  observed abundance of solid water and together with the derived
  HDO/\hho\ abundance ratios of $\sim$10$^{-3}$ suggests that the origin of the
  observed water is evaporation of grain mantles.
  This idea is confirmed in the case of AFGL 2591 by interferometer observations
  of the HDO $1_{10}$--$1_{11}$, \hhoe\ $3_{13}$--$2_{20}$ and \soo\ 
  $12_{0,12}$--$11_{1,11}$ lines, which reveal compact (\O$\sim$800~AU)
  emission with a systematic velocity gradient. This size is similar to that of
  the 1.3~mm continuum towards AFGL 2591, from which we estimate a mass of
  $\approx$0.8~\msol, or $\sim$5\% of the mass of the central star. We speculate
  that we may be observing a circumstellar disk in an almost face-on
  orientation.

\keywords{ISM: molecules -- Molecular processes -- Stars: formation -- Astrochemistry}
}
\maketitle

\section{Introduction}
\label{s:intro}

Water is a cornerstone molecule in the oxygen chemistry of dense interstellar
clouds and a major coolant of warm molecular gas\footnote{This paper uses the word
  `water' to denote the chemical species, and the notations \hho, \hhoe\ and HDO
  to denote specific isotopologues.}.
In the surroundings of embedded protostars, water can be formed by three very
different mechanisms.
In cold ($\sim$10~K) protostellar envelopes, water may be formed in the gas
phase by ion-molecule chemistry, through dissociative recombination of \hhhop.
Simultaneously, on the surfaces of cold dust grains, O and H atoms may combine
to form water-rich ice mantles. These mantles will evaporate when the grains are
heated to $\sim$100~K, either by protostellar radiation or by grain sputtering
in outflow shocks.
Third, in gas with temperatures $\gtsim$250~K, reactions of O and OH with \hh\ 
drive all gas-phase oxygen into water. Such high temperatures may occur very
close to the star due to radiation, or further out in outflow shocks.
The water molecule thus offers the opportunity to study the relative
importance of each of these types of chemistry in the protostellar environment.

There has been considerable controversy about the water abundance around
high-mass protostars.
Observations of the \hho\ 6~\mic\ bending mode with ISO-SWS have revealed
abundant water (\hho/\hh\ $\sim$10$^{-5}$--10$^{-4}$) in absorption toward several
high-mass protostars \citep{boonman:h2o}. The absorption data do not tell us the
location of the \hho\ along the line of sight, except that the high excitation
temperatures ($\sim$300--500~K) imply an origin in warm gas.
In contrast, observations of the o-\hho\ ground state line at 557~GHz with SWAS
of the same sources indicate much lower abundances (\hho/\hh\ 
$\sim$10$^{-7}$--10$^{-6}$; \citealt{snell:swas}).  The narrow line width
indicates an origin in the envelopes rather than the outflows of the sources,
but the data have too low angular resolution (several arcminutes) for more
detailed statements.
\new{\citet{boonman:models} performed a simultaneous analysis of ISO-SWS,
ISO-LWS and SWAS data and inferred a water abundance jump in the inner
envelope by four orders of magnitude for several high-mass YSOs.}

\begin{table*}[t]
    \caption{Source sample.}
    \label{tab:samp}
    \begin{tabular}[lccccc]{lccccc}
\hline \hline
Source        & R.A.\ (1950)& Dec.\ (1950) & $L$ & $d$ & $N$(\hh)$^a$ \\
 &(h m s) & (\degr\ \arcmin\ \arcsec) & ($10^{4}$~\lsol) & (kpc) & ($10^{23}$~\scm)\\
\hline
W3 IRS5       & 02 21 53.1 & +61 52 20 & 17 & 2.2 & 2.3 \\
AFGL 490      & 03 23 38.9 & +58 36 33 & 0.2 & 1  & 2.0 \\
W33A          & 18 11 43.7 &--17 53 02 & 10 & 4   & 6.2 \\
AFGL 2136     & 18 19 36.6 &--13 31 40 &  7 & 2   & 1.2 \\
AFGL 2591     & 20 27 35.8 & +40 01 14 &  2 & 1   & 2.3 \\
S140 IRS1     & 22 17 41.1 & +63 03 42 &  2 & 0.9 & 1.4 \\
NGC 7538 IRS1 & 23 11 36.7 & +61 11 51 & 13 & 2.8 & 6.5 \\
NGC 7538 IRS9 & 23 11 52.8 & +61 10 59 &  4 & 2.8 & 3.3 \\
\hline
    \end{tabular}

\new{$^a$: Column density in a 15$''$ beam.}

\end{table*}

Locating the water around protostars requires observations at high spatial and
spectral resolution, which presently can only be done from the ground. Most
ground-based observations of \hho\ have targeted the maser lines at 22 and
183~GHz (e.g., \citealt{cernicharo:h2o}). However, the anomalous excitation of
these lines makes it hard to derive column densities from such data, which may in
any case not be representative of the surrounding region.
The only thermal water lines that can be studied from the ground are the
$3_{13}$--$2_{20}$ line of \hhoe\ at 203~GHz (\citealt{phillips:h2o18};
\citealt{jacq:h2o18}), and several HDO lines.
\new{These lines were used by \citet{gensheimer:h2o} and \citet{helmich:hdo}
  to estimate envelope-averaged abundances of \hho\ and HDO around
  several young high-mass stars. }
\new{Advances in sensitivity and resolution allow us to consider
  lower-luminosity objects closer than the Sun than before, and also
  enable us to study abundance variations with position in the
  envelope. }
This paper presents \new{new } observations of these lines toward sources that have been
studied previously with ISO and SWAS, including the first published interferometric
observations of the \hhoe\ line (and in fact of any non-masing water line). 
\cut{The goal is to find the location of the water seen with the space observatories.}
The sources are eight deeply embedded high-mass stars, with
luminosities of 2$\times$10$^3$ -- 2$\times$10$^5$~\lsol, distances of
1 -- 4~kpc, 
and \hh\ column densities of 1 -- 7$\times$10$^{23}$~\scm, as listed in
Table~\ref{tab:samp}. Single-dish mapping of dust continuum and molecular line
emission at \smm\ wavelengths indicates envelope masses of 30 -- 1100~\msol\ 
within radii of 0.09 -- 0.36~pc \citep{vdtak:massive}.
The sources drive powerful outflows as revealed by mid-infrared and
millimeter-wave observations of CO and HCO$^+$ (\citealt{mitchell:episodic};
\citealt{hasegawa:outflow}). 
The unique aspect of this source sample is its high mid-infrared brightness,
which allows us to compare its \smm\ emission with solid state data for the
chemistry, and with rovibrational absorption lines for the geometry.

This paper is organized as follows. Section~\ref{s:obs} describes the
observations, and Section~\ref{s:res} their direct results.
Section~\ref{s:radtrans} describes modeling of the data with a radiative
transfer program. Section~\ref{s:disc} discusses the results of the observations
and the models in the context of a disk/outflow geometry for these sources.
Section~\ref{s:conc} concludes the paper with an outlook toward future
opportunities in this field.

\section{Observations}
\label{s:obs}

Table~\ref{t:lines} summarizes spectroscopic parameters of the
observed lines, and gives the relevant telescope and its FWHM beam
size at that frequency. 
With $E_{\rm up}\approx 200$~K, the $3_{13}$ -- $2_{20}$ line is the
lowest-lying transition of \hhoe\ that can be observed from the
ground.  We use this line to measure the abundance of \hho\ in the
warm inner envelopes of the sources.
The HDO lines cover the range of excitation energies from 20 to 200~K,
and are used to constrain the excitation and chemical state of the
gas, in particular its deuterium fractionation.
The \soo\ and \meth\ lines have comparable excitation requirements,
and are used to measure the effects of shock chemistry (\soo ) and
ice evaporation (\meth ).
The difference in Einstein $A-$coefficients of the lines is mostly due
to the $\nu^3$ dependence: all the lines have transition dipole
moments of a few Debye.

\begin{table*}
\caption{Observed transitions.}
\label{t:lines}
\begin{tabular}{lcrrccc}

\hline
\hline

Species & Transition & Frequency & $E_{\rm up}$ & $A_{ul}$ & Telescope & Beam \\
        & $J_{K_pK_o}$ & MHz     & K            & s$^{-1}$ &           & $''$ \\

\hline

HDO     & $1_{10}$--$1_{11}$ &  80578.3 &  47   & 1.3$\times$10$^{-6}$ & IRAM 30m & 30 \\
HDO     & $3_{12}$--$2_{21}$ & 225896.7 & 168   & 1.3$\times$10$^{-5}$ & IRAM 30m & 11 \\
HDO     & $2_{11}$--$2_{12}$ & 241561.6 &  95   & 1.2$\times$10$^{-5}$ & JCMT 15m & 21 \\
HDO     & $1_{01}$--$0_{00}$ & 464924.5 &  22   & 1.7$\times$10$^{-4}$ & JCMT 15m & 12 \\

\hhoe\  & $3_{13}$--$2_{20}$ & 203407.5 & 204   & 4.9$\times$10$^{-6}$ & IRAM 30m & 12 \\

\soo\   & $12_{0,12}$--$11_{1,11}$ &203391.6 & 70 & 8.1$\times$10$^{-5}$ & IRAM 30m & 12 \\

\meth\  & $5_{-1}$--$4_0$ E  &  84521.2 &  40   & 2.0$\times$10$^{-6}$ & IRAM 30m & 30 \\

\hline
\end{tabular}
\end{table*}

\subsection{Single-dish observations}
\label{s:sd-obs}

Observations of lines of \hhoe, HDO, \soo\ and \meth\ in the 80 -- 225~GHz
range were made with the 30-m telescope of the Institut de Radio
Astronomie Millim\'etrique (IRAM)\footnote{IRAM is an international
  institute for research in millimeter astronomy, co-funded by the
  Centre National de la Recherche Scientifique (France), the Max
  Planck Gesellschaft (Germany) and the Instituto Geografico Nacional
  (Spain).} on Pico Veleta, Spain, in May 2003. The front ends were
the facility receivers A100, B100, A230 and B230, and the backend was
the Versatile Spectral Assembly (VESPA) autocorrelator. The five lines
were observed simultaneously with a spectral resolution of
0.1 -- 0.3~\kms.  Integration times are 60 -- 180 minutes (on+off) using
double beam switching with a throw of 180$''$. System temperatures
were 100 -- 150~K at 3~mm and 300 -- 600~K at 1.3~mm wavelength.
Data were calibrated onto \tmb\ scale by multiplying by $\eta_f /
\eta_b$, where the forward efficiency $\eta_f$ is 0.95 at 3~mm and
0.91 at 1.3~mm, and the main beam efficiency $\eta_b$ is 0.78 at
3~mm and 0.57 at 1.3~mm wavelength.
\new{The spectra have noise levels per 0.25~\kms\ channel of \tmb =10 -- 15~mK
  at 80~GHz and 20 -- 30~mK at 215~GHz.}

Additional observations of HDO lines at 225, 241 and 464~GHz 
toward selected sources were carried
out with the James Clerk Maxwell Telescope (JCMT)\footnote{The JCMT is
  operated by the Joint Astronomy Centre, on behalf of the Particle
  Physics and Astronomy Research Council of the United Kingdom, the
  Netherlands Organization for Scientific Research and the National
  Research Council of Canada.} on Mauna Kea, Hawaii, in 1995 -- 1997.
These data were taken as part of a spectral line survey program, and have
a lower spectral resolution and signal to noise ratio than the IRAM spectra.
The facility receivers A2 and C2 were used as front ends and the Dutch
Autocorrelation Spectrometer (DAS) as back end. The JCMT has a main
beam efficiency of 0.65 at 1.3~mm and 0.53 at 0.6~mm wavelength.
Integration times are 30 minutes at 241~GHz and 40 minutes at 464~GHz,
resulting in rms noise levels in \tmb\ per 625~kHz channel of
$\approx$40~mK at 241~GHz and $\approx$200~mK at 464~GHz.

All single-dish spectra have been reduced with the CLASS package, developed at
IRAM\footnote{\tt http://www.iram.fr/IRAMFR/GILDAS/}. Linear baselines
were subtracted and the spectra were smoothed once and calibrated onto
\tmb\ scale. We estimate a calibration uncertainty of 30\% for the
80~GHz IRAM and 240~GHz JCMT data, and of 50\% for the 230~GHz IRAM
and 460~GHz JCMT data, due to higher atmospheric opacity.
\new{The estimated pointing uncertainty for all single-dish data is 3$''$ rms.}

\subsection{Interferometric observations}
\label{s:pdb-obs}

The IRAM interferometer on Plateau de Bure (France) consists
of six 15-m antennas on North-South and East-West baselines.
We used this instrument to map the HDO $1_{10}$--$1_{11}$,
\hhoe\ $3_{13}$--$2_{20}$ and \soo\ $12_{0,12}$--$11_{1,11}$
lines and continuum at 80.6~and 204.9~GHz toward AFGL 2591.
Due to tuning problems, only five antennas took 1.3~mm data in
$C$--configuration on December 6, 2003; these problems were
solved before the $D$--array observations on May 15 -- 16,
2004. 
The correlator was configured to produce `narrow' 80~MHz and `broad' 160~MHz
windows, with one of each centered on the HDO and \hhoe\ lines. The number of
channels is 128 per window. The \soo\ line falls in the 160~MHz window of the
\hhoe\ line. The continuum bandwidth is 640~MHz at 81~GHz and twice as much at
205~GHz, where tuning is double side band.
Antenna gains and phases were monitored by observing
2013+370 and 2005+403 for 2~minutes every 20~minutes.
The total observing time was 13.2~hr of good weather in
$C$--array and 10~hr of excellent weather in $D$--array.
The combined dataset has baselines ranging from the
antenna shadowing limit out to 309~m, corresponding to an
angular scale of 2.2$''$ at 81~GHz and 0.85$''$ at 205~GHz.
Data reduction was performed at the IRAM headquarters in
Grenoble, using the GILDAS software.
Bandpass was checked on 3C273 and 2145+067.
Flux calibration was performed on MWC 349, assuming
$S_\nu$=1.0~Jy at 81~GHz and 2.0~Jy at 205~GHz.

\section{Results}
\label{s:res}

\begin{figure}[tb]
  \begin{center}
\includegraphics[width=6cm,angle=0]{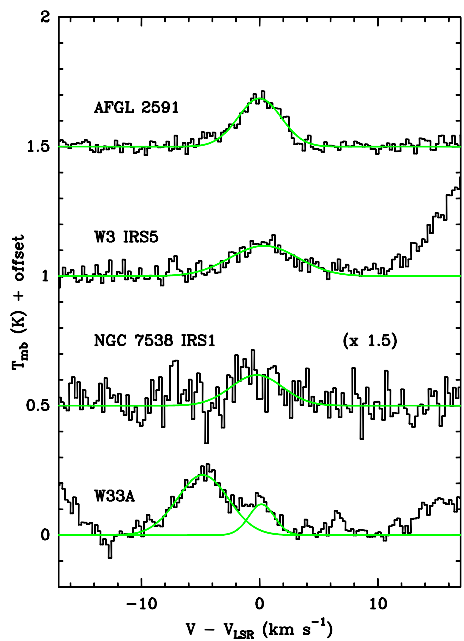}
\caption{Spectra of the \hhoe\ $3_{13}$--$2_{20}$ transition, obtained
  with the IRAM 30m telescope. The line at $V$=$+$16~\kms\
  is the \soo\ $12_{0,12}$--$11_{1,11}$ transition; the lines in
  the W33A spectrum at $V$=$-$5 and $-$17~\kms\ are due to CH$_3$OCH$_3$.}
\label{f:30m}
\end{center}
\end{figure}

\begin{figure}[tb]
  \begin{center}
\includegraphics[width=7cm,angle=0]{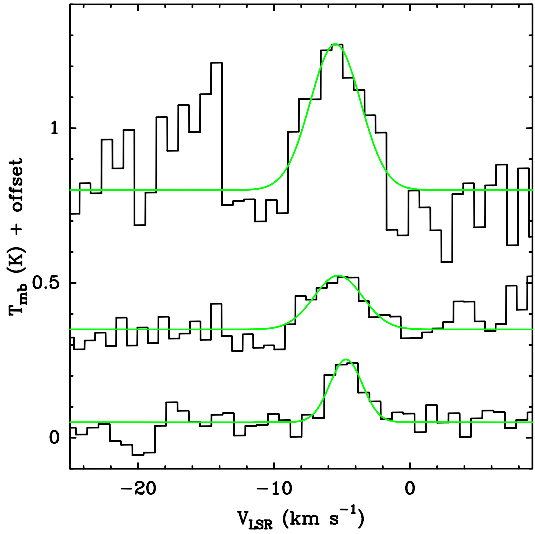}
\caption{Observations of HDO lines toward AFGL 2591 with the JCMT. Top to
  bottom: 465 GHz, 241 GHz, 225 GHz lines. The bottom two spectra have been
  multiplied by 2 and all spectra are vertically offset for clarity. The
feature at \vlsr=--15~\kms\ in the upper spectrum is due to the \meth\
$18_{8,10}$--$18_{9,10}$~E line.}
\label{f:jcmt}
\end{center}
\end{figure}

\begin{table*}
\caption{Line fluxes (K \kms) or 1$\sigma$ upper limits (mK) observed with the IRAM 30m.
         Numbers in brackets are uncertainties in units of the last
         decimal place. \new{The errors do not take calibration into
         account but only spectral noise. } }
\label{t:flux}
\begin{tabular}{lcrrcr}

\hline
\hline

Source  & HDO & \meth\ & \hhoe\ & \soo\ & HDO \\
  & $1_{10}$--$1_{11}$ & $5_{-1}$--$4_0$~E & $3_{13}$--$2_{20}$ & $12_{0,12}$--$11_{1,11}$ & $3_{12}$--$2_{21}$ \\

\hline

W3 IRS5       & 0.14(2) & 0.17(2) & 0.84(7) & 19.44(8) & $<$36 \\
AFGL 490      & $<$13   & 0.39(2) & $<$24   & $<$24    & $<$31 \\
W33A          & 0.66(3) & 7.69(3) & 0.46(2) & 4.73(2)  & 4.06(6) \\
AFGL 2136     & $<$9    & 0.50(1) & $<$21   & 0.85(3)  & $<$27 \\
AFGL 2591     & 0.15(1) & 1.51(1) & 0.86(3) & 4.01(3)  & 0.59(3) \\
S140 IRS1     & $<$10   & 1.41(1) & $<$22   & 2.26(3)  & $<$23 \\
NGC 7538 IRS1 & 0.26(3) & 2.64(2) & 0.43(9) & 2.09(8)  & 1.57(8) \\
NGC 7538 IRS9 & $<$12   & 2.07(2) & $<$22   & 0.62(4)  & $<$29 \\

\hline
\end{tabular}
\end{table*}

\begin{table*}
\caption{Widths (FWHM in \kms) of the lines observed with the IRAM 30m. 
         Numbers in brackets are uncertainties in units of the last decimal place.}
\label{t:width}
\begin{tabular}{lcrrcr}

\hline
\hline

Source  & HDO & \meth\ & \hhoe\ & \soo\ & HDO \\
 & $1_{10}$--$1_{11}$ & $5_{-1}$--$4_0$ E & $3_{13}$--$2_{20}$ & $12_{0,12}$--$11_{1,11}$ & $3_{12}$--$2_{21}$ \\

\hline

W3 IRS5       & 3.2(5) & 2.7(4)      & 6.8(10) & 6.67(4)$^a$ & ... \\
AFGL 490      & ...    & 2.5(2)$^a$  &   ...   &   ...       & ... \\
W33A          & 4.6(2) & 4.67(2)$^a$ & 3.8(1)  & 6.1(1)$^a$  & 5.06(9)$^a$ \\
AFGL 2136     & ...    & 2.79(7)$^a$ &   ...   & 4.2(2)      & ... \\
AFGL 2591     & 3.3(4) & 2.87(3)$^a$ & 4.3(2)  & 5.33(5)     & 3.2(2) \\
S140 IRS1     & ...    & 2.68(3)$^a$ &   ...   & 2.67(4)$^a$ & ... \\
NGC 7538 IRS1 & 3.6(5) & 3.21(3)$^a$ & 5.2(16) & 6.1(3)      & 3.8(2) \\
NGC 7538 IRS9 & ...    & 2.94(3)$^a$ &   ...   & 5.3(4)      & ... \\

\hline
\end{tabular}
\medskip

{\scriptsize a}: Line core only; also wings visible

\end{table*}

\begin{table}
\caption{JCMT observations of HDO.}
\label{t:jcmt}
\begin{tabular}{lcccc}

\hline
\hline

Line & $\int$\tmb\textit{dV} & \vlsr & \dv  & \tmb \\
     & \kkms                 & \kms  & \kms & K \\
\hline

\multicolumn{5}{c}{AFGL 2591} \\

$3_{12}$--$2_{21}$ & 0.308(48) & --4.70(22) & 2.83(52) & 0.10 \\
$2_{11}$--$2_{12}$ & 0.394(80) & --5.27(45) & 4.24(88) & 0.09 \\
$1_{01}$--$0_{00}$ & 2.19(41)  & --5.48(43) & 4.35(81) & 0.47 \\

\multicolumn{5}{c}{NGC 7538 IRS1} \\

$3_{12}$--$2_{21}$ & 0.74(12) & --58.43(28) & 3.57(62) & 0.20 \\

\hline
\end{tabular}

\end{table}

\subsection{Single-dish spectra}

With the IRAM 30m, we have detected \hhoe\ in four objects
(Fig.~\ref{f:30m}). The HDO 80~GHz line is detected in the same four
objects, and the 225~GHz line in three of them. The \meth\ line is
seen in all eight sources, and the \soo\ line in all but one.
Tables~\ref{t:flux} and~\ref{t:width} list the integrated intensities
and widths of the lines, obtained through Gaussian fits to the
profiles. \new{Note that for strong lines, calibration dominates the
  uncertainty on the line flux, while spectral noise dominates for
  weak lines. }
The spectra also show a few unexpected lines. In W33A, the
$4_{04}$--$3_{03}$ line of formamide (NH$_2$CHO) at 84542.4~MHz was
detected with \tmb\ = 42~mK and \dv\ = 5.6~\kms. Next to the
\hhoe\ line, the $3_{30}$--$2_{21}$, $3_{31}$--$2_{21}$ and
$3_{30}$--$2_{20}$ lines of dimethyl ether (CH$_3$OCH$_3$) at 203420,
203410 and 203384~MHz are detected with \tmb\ = 0.21, 0.23 and 0.25~K
and \dv\ = 4.6, 4.4 and 4.9~\kms. The $3_{30}$--$2_{20}$ line of
CH$_3$OCH$_3$ is also detected toward NGC 7538 IRS1, with \tmb\ =
0.12~K and \dv\ = 3.7~\kms.

With the JCMT, we have detected three HDO lines in AFGL 2591, and one
in NGC 7538 IRS1 (Fig.~\ref{f:jcmt}; Table~\ref{t:jcmt}). Upper
limits \new{(1$\sigma$) } of \tmb = 0.24, 0.35 and 0.17~K on 0.625~MHz channels were
obtained for the 464~GHz line toward W33A, AFGL 2136 and S140 IRS1.
For the 225~GHz line, upper limits of \tmb\ = 38~mK were found for
S140 IRS1 and NGC 7538 IRS9. 
\citet{helmich:hdo} report a tentative detection of the 464~GHz line
toward W3 IRS5 and upper limits on the 225 and 241 GHz lines;
\citet{schreyer:gl490} set an upper limit to the 464~GHz emission from AFGL 490.
The JCMT spectra of HDO have a lower signal to noise ratio and spectral
resolution than the IRAM 30m spectra, and the line positions and widths in
Table~\ref{t:jcmt} are too uncertain to extract kinematic information.

The central velocities and the widths of the HDO lines in the 30m spectra (Table~\ref{t:width}) 
are consistent with the values for the molecular envelopes of these
objects, derived from C$^{17}$O and C$^{34}$S spectra \citep{vdtak:massive}.
In contrast, the \hhoe\ lines are 30 -- 90\% broader than the HDO lines in the
same sources (Table~\ref{t:width}), which may be an indication
that part of the \hhoe\ emission arises in the molecular outflows of these
sources.  Only in W33A, the fitted width of the \hhoe\ line is smaller than that
of the HDO lines, but for this source, the \hhoe\ line is blended with other
lines (see Fig.~\ref{f:30m}), making its width hard to measure.

\begin{figure}[tb]
  \begin{center}
\includegraphics[width=7cm,angle=0]{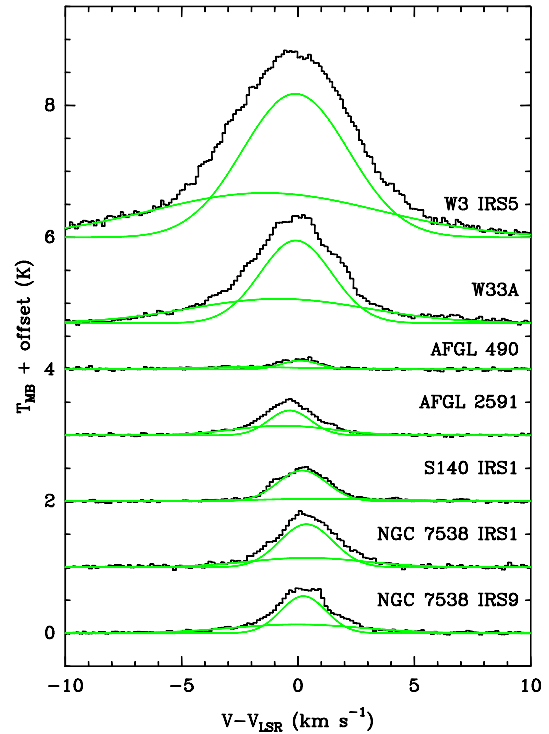}
\caption{Line profiles of \meth, observed with the IRAM 30m, with
  two-component fits overplotted. For W3 IRS5, the \soo\ line is
  plotted instead of the \meth\ line.}
\label{f:meth}
\end{center}
\end{figure}

Evidence for a contribution to the observed emission by outflows is
even more pronounced in the \soo\ and \meth\ spectra, which have
higher signal-to-noise than those of HDO and \hhoe.
The profiles of \soo\ in three sources and of \meth\ in seven show
low-level emission at high velocities (Table~\ref{t:width}).
We have fitted these profiles with the sum of two independent
Gaussians: a narrow one corresponding to the `envelope' component seen
in C$^{17}$O and C$^{34}$S, and a broader one (Fig.~\ref{f:meth})
which we attribute to the outflow.
We find widths for the broad components between 4.7~\kms\ in AFGL
2591 and 12.8~\kms\ in W3 IRS5; the source-averaged width is 7.1~\kms. 
The broad component is blueshifted from the narrow one in all
sources except S140 IRS1, which ties in with the tendency for
blueshifted mid-infrared absorption of CO \citep{mitchell:episodic},
which also arises in outflows.
The fraction of the line flux carried by the broad \new{\meth\ } component ranges from 16\% in
S140 IRS1 to 58\% in AFGL 2591, and is 39\% on average.  These fractions are
comparable with the values of 10 -- 50\% found for CS, SO and \soo\ in these
sources \citep{vdtak:sulphur} \new{which may depend on excitation
  energy or beam size }.
Previous \meth\ spectra of these sources which had lower signal to noise ratios
and lower spectral resolution did not show line wings, although they did
show wings for two other, similar objects \citep{vdtak:meth}.

\new{The line profiles of \hhoe\ and HDO do not have high enough signal to noise
ratios to perform two-component fits. 
Therefore we assume that the outflow contribution estimated for CS, SO, \soo\ 
and \meth\ also holds for \hhoe\ and HDO, which have similar excitation
requirements and optical depths.
However, instead of a double Gaussian fit, the outflow contribution
may be estimated as the line flux at high velocities only. The result
is somewhat less than half of that from a double Gaussian fit,
or $\sim$10 -- 20\%. This value is probably consistent with the fraction of
$\approx$50\% estimated for the o-H$_2^{16}$O ground state line
observed with SWAS (\citealt{snell:swas}; \citealt{boonman:models}),
given the different excitation requirements.
}

The presence of significant amounts of \meth\ in the outflow raises
the question of its desorption from grain mantles. 
Toward young low-mass protostars, the methanol emission of transitions with low
energy levels is dominated by outflow gas \citep{bachiller:l1157}, indicating
that grain mantle desorption by shocks is at work. Higher excitation methanol
lines are broad toward some low-mass protostars and
  narrow towards others, indicating that shocks or thermal desorption
may dominate in the warmer regions of the envelope (\citealt{maret:ch3oh};
\citealt{jorgensen:methanol}).
The present data show that for high-mass protostars, radiation and shocks both
have a relevant role in releasing ice mantles from dust grains \new{too}.

\subsection{Excitation of HDO}
\label{s:txc}

For the sources where several HDO lines are detected, we have
estimated the excitation temperature using rotation diagrams.
This method, described in detail by \citet{blake:orion} and
\citet{helmich:w3}, assumes that the lines are optically thin and
describes the molecular excitation by a single temperature, the
`rotation temperature'.
However, this temperature is only meaningful if all the data refer to
the same physical volume.
The beam sizes of our observations range from 12 to 30$''$, and it is important
to consider the effect of beam dilution.
Indeed, if the HDO emission from W33A, AFGL 2591 and NGC 7538 IRS1 were extended
on scales as large as 30$''$, the upper-state column densities of the
higher-excitation lines would be larger than those of the low-excitation lines,
implying an infinite \new{or negative } excitation temperature. 
\cut{To avoid this unphysical situation, }
\new{Since such non-thermal excitation is unlikely, }
the data must be corrected for the effect of a finite source size.

The size of the HDO emission can be estimated for AFGL 2591 and NGC
7538 IRS1, where the 225~GHz line has been measured both with the IRAM
30m and the JCMT. The emission is about twice as bright in the IRAM
30m beam, suggesting a compact source size. Therefore we assume a
source size of 12$''$ for the HDO in W33A, AFGL 2591 and NGC 7538
IRS1, and correct the 80~GHz line fluxes upward by (30/12)$^2$.
This factor is much larger than any plausible optical depth effect on
the 80~GHz line, especially since the 464~GHz line, which lies lower
in excitation, is expected to have a larger optical depth.
Statistical equilibrium calculations indeed indicate optical depths of
$\sim$10$^{-2}$ for the excitation temperatures and brightness levels
of HDO found here.

Table~\ref{t:txc} lists the assumed sizes for all sources where HDO
has been detected, and the resulting HDO excitation temperatures.
The data do not rule out source sizes $<$12$''$, and the assumed size may be
regarded as an upper limit. Smaller source sizes would influence all lines
equally, though, and not change the excitation temperature estimates.
In the case of W3 IRS5, the only firm detection of HDO is the 80~GHz line. 
The observational limits on the 225, 241 and 464~GHz lines indicate
\txc=85--115~K, but do not constrain the source size.

\begin{table}
\caption{Sizes and excitation temperatures of the HDO emission,
  \new{derived from the single-dish observations.} }
\label{t:txc}
\begin{tabular}{lcc}

\hline
\hline

Source        & Size & \txc \\
              & $''$ & K \\

\hline

W3 IRS5       & ... & 85 -- 115 \\
W33A          & 12     & 110$\pm$58 \\
AFGL 2591     & 12     & 117$\pm$57 \\
NGC 7538 IRS1 & 12     & 108$\pm$56 \\

\hline

\end{tabular}
\end{table}


The excitation temperatures found for HDO may be used as a first clue to its
chemical origin by comparison with \soo\ (a product of shock chemistry), \meth\ 
(a product of ice evaporation), and \cchh\ (a product of hot gas-phase
reactions).
The excitation temperatures of HDO are similar to the values of
50 -- 200~K derived for \meth\ \citep{vdtak:meth} and \soo\ 
\citep{vdtak:sulphur}, measured in \smm\ emission in 14 -- 18$''$ beams.
These excitation temperatures are lower limits to the kinetic temperature of the
emitting gas, but this \cut{effect} probably has little effect on the comparison with HDO
since the molecules have similar dipole moments.
The excitation temperatures are much lower than the values of 500 -- 1000~K
measured in mid-infrared absorption of \cchh\ \citep{lahuis:hcn} as expected for two
reasons. First, the \cchh\ molecule does not have a permanent dipole moment, and
its excitation temperature directly reflects the kinetic temperature of its
surroundings. Second, \cchh\ is seen in absorption, which in these centrally
condensed envelopes preferentially probes smaller radii, where the temperature
is higher. 
The excitation temperatures of \new{HCN 14~\mic\ and } \hho\ 6.2~\mic\ absorption
of 300 -- 500~K for these sources \citep{boonman:h2o} are between the values for
HDO emission and \cchh\ absorption, suggesting that the two effects contribute
about equally.

\begin{figure}[tb]
  \begin{center}
\includegraphics[width=5cm,angle=0]{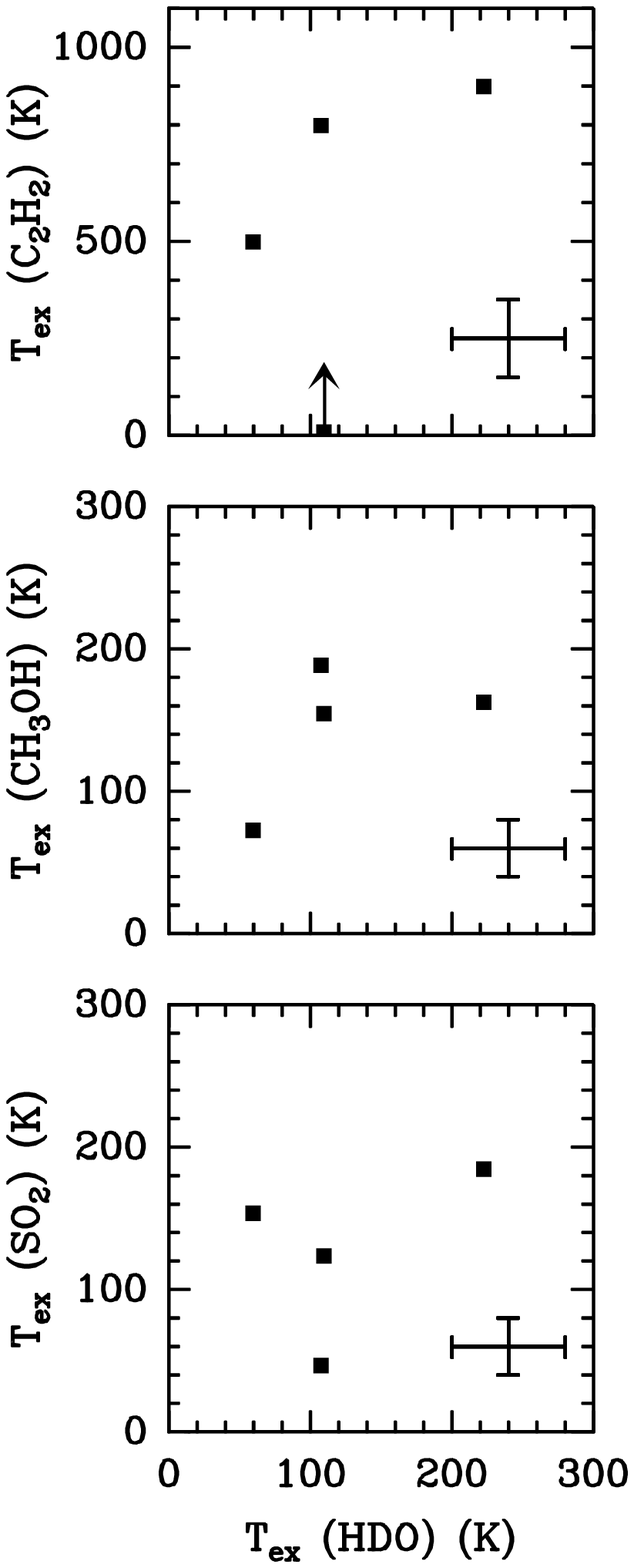}
\caption{Excitation temperatures of HDO plotted versus values of \soo\
  (bottom), \meth\ (middle) and \cchh\ (top). The low \txc\ of \cchh\ in W33A 
  may be affected by a high continuum optical depth at 14~\mic. }
\label{f:txc}
\end{center}
\end{figure}

The HDO, \meth, \soo\ and \cchh\ in these objects are thus located in warm
(several 100~K) gas, such as found in the inner envelopes and outflow shocks of protostars.
The exact temperature is hard to estimate because of the above caveats,
so that we have searched for trends instead.
Even here, the data appear inconclusive (Fig.~\ref{f:txc}).
Further discussion of the origin of the HDO is deferred until after the
radiative transfer model in \S~\ref{s:disc}.

\subsection{Column densities of HDO and \hho }
\label{s:cold}

\begin{figure}[tb]
  \begin{center}
\includegraphics[width=5cm,angle=-90]{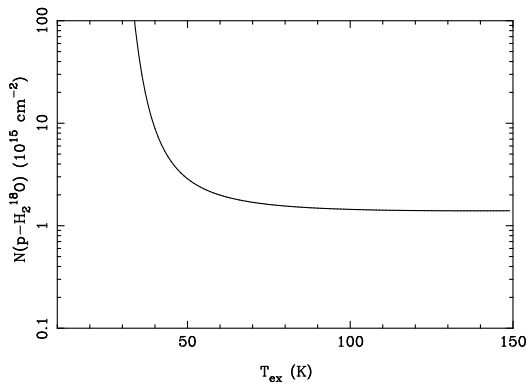}
\caption{Sensitivity of the derived $p-$\hhoe\ column density to the
  adopted excitation temperatures, for a line flux of 1.0~K\,\kms.}
\label{f:nh2o}
\end{center}
\end{figure}

Knowing the excitation conditions, we derive molecular column
densities from the observed line strengths. For HDO, these follow
directly from the rotation diagrams. For \hho, these come from the
\hhoe\ data, assuming optically thin \hhoe\ emission with the same
excitation temperature as the HDO in that source.  We use an oxygen
isotope ratio of $^{16}$O/$^{18}$O=500 and assume an ortho/para ratio
of 3 for \hho, as expected for warm gas.
The resulting column densities (Table~\ref{t:cold}) are uncertain by a
factor of $\approx$2, mainly through the uncertain excitation temperature.
The sensitivity of $N$(\hho) to \txc\ is such that increasing \txc\ 
from 110 to 220~K increases the derived column density just slightly,
while decreasing \txc\ from 110 to 60~K almost doubles it and further
lowering \txc\ leads to implausibly high \hho\ column densities
(Figure~\ref{f:nh2o}).
%
%
In addition, if the source size is smaller than the 12$''$ assumed above, the
column density estimates would increase.

For the four sources where HDO and \hhoe\ were not detected, the noise
levels of the spectra imply limits on the column densities. Assuming
\txc=110~K and \dv=3.0~\kms, the 3$\sigma$ limits \new{in an 11$''$
  beam } are
$N$(HDO)$<$6.5$\times$10$^{13}$~\scm\ and
$N$(\hho)$<$6.4$\times$10$^{16}$~\scm, which for both species is a factor
of $\approx$2 below the weakest detection.
In the cases of AFGL 490, S140 IRS1 and NGC 7538 IRS9, the non-detection
of HDO and \hhoe\ emission ties in with non-detections of \hho\ 6.2~\mic\ 
absorption (\citealt{boonman:h2o}; \citealt{schreyer:gl490}) and low
column densities of warm \hh\ as traced by $^{13}$CO 4.7~\mic\ absorption
(\citealt{mitchell:hot+cold}; \citealt{mitchell:gl490}).
\cut{For AFGL 490, \hho\ 6.2~\mic\ data are not available, but the column
density of warm \hho and CO is also low \citep{mitchell:gl490}.}
In contrast, AFGL 2136 does have high \hho\ and \hh\ column densities
measured in mid-infrared absorption, but this gas must be compact,
since the \hh\ column density from \smm\ data is low \citep{vdtak:sulphur}.

\new{The derived molecular column densities may be used as a clue to
  the source geometry by comparing them with values measured in
  mid-infrared absorption. }
The \hho\ column densities in Table~\ref{t:cold} are consistent with the values
from ISO 6.2~\mic\ absorption \citep{boonman:h2o} to within factors of a few.
This situation is similar to that for CO and dust, where \smm\ data indicate
column densities $\sim$3--5$\times$ higher than mid-infrared data \citep{vdtak:massive}. 
\new{Therefore, if these species are spherically distributed around the
  central star, this region must be extended on the
scales of the single-dish beams } (diameter $\gtsim$1$''$, corresponding to
$\gtsim$2000~AU at $d$=2~kpc),
\cut{although this conclusion depends on the geometry of the emitting
  region.}
\new{Alternatively, these molecules do not have spherically symmetric
  distributions. }
In contrast, for \soo\ and HCN, mid-infrared absorption lines indicate
$\sim$100$\times$ higher column densities than \smm\ emission lines
(\citealt{keane:so2}; \citealt{boonman:hcn}). These molecules must have
distributions as compact as $\ltsim$0\farcs1 ($\ltsim$200~AU)
which may or may not be spherical.
Additional constraints on the source geometry come from the
  interferometer data (\S~\ref{s:bure_lines}).

\begin{table*}
\caption{Column densities of HDO and \hho\ in an 11$''$ beam,
  \new{derived from the single-dish observations.} }
\label{t:cold}
\begin{tabular}{lccccr}

\hline
\hline

Source        & HDO & \hho           & \hho $^a$ & $\bar{T}$ $^d$ & HDO/\hho \\
   & 10$^{14}$ \scm & 10$^{17}$ \scm & 10$^{17}$ \scm & K        & 10$^{-4}$ \\

\hline

W3 IRS5 & 0.3 -- 0.6$^b$ & 2.6 -- 5.0$^c$ & 3    & 33 &  1 \\
W33A          & 6.9      & 1.4            & $<$8 & 20 & 70 \\
AFGL 2591     & 2.3      & 3.0            & 4    & 28 &  8 \\
NGC 7538 IRS1 & 2.7      & 1.3            & $<$5 & 25 & 30 \\

\hline

\end{tabular}

\medskip

{\scriptsize a}: From ISO 6.2~\mic\ absorption \citep{boonman:h2o} in a pencil beam.

{\scriptsize b}: Values for source sizes of 30$''$ and 12$''$.

{\scriptsize c}: Values for \txc=110 and 60~K.

{\scriptsize d}: Mass-weighted envelope temperature from \citet{vdtak:massive}.

\end{table*}

The $N$(HDO) / $N$(\hho) ratios (Table~\ref{t:cold}, column~6) are consistent
with the limits on solid-state HDO/\hho\ obtained for these same sources
\citep{dartois:hdo},
and similar to the values measured for `hot core' type
regions (\citealt{jacq:hdo}; \citealt{gensheimer:h2o}).
The ratios correspond to enhancements of the HDO/\hho\ ratio over the elemental
abundance ratio (D/H=1.5$\times$10$^{-5}$; \citealt{linsky:d}) of 7 for W3 IRS5,
50 for AFGL 2591, 200 for NGC 7538 IRS1 and 470 for W33A.
The enhancement level shows a correlation with the mass-weighted average
envelope temperature $\bar{T}$ of these sources \citep{vdtak:massive}, listed in
the fifth column of Table~\ref{t:cold}, in the sense that colder sources have
higher HDO/\hho\ ratios.  The decrease of the HDO enhancement from 470 in W33A
($\bar{T}$=20~K) to 7 in W3 IRS5 ($\bar{T}$=33~K) is the combined result of a
decrease in $N$(HDO) by a factor of 16 and an increase in $N$(\hho) by a factor
of 4.
%
%
The HDO/H$_2$O ratios in Table~\ref{t:cold} are 10 -- 100$\times$ higher than
the equilibrium value at the estimated gas temperatures ($\sim$few 100~K), which
suggests that the HDO (and H$_2$O) molecules are a remnant from an earlier,
colder phase of the protostars (like in the case of low-mass protostars; see
\S~\ref{ss:lmpo}).  A natural explanation is that HDO (and H$_2$O) molecules are
sublimated from the grain mantles in the warm region where the dust temperature
exceeds 100~K, the ice sublimation temperature. In this scenario, the coldest
sources would be also the youngest, where gas-phase reactions occurring in the
warm region containing the sublimated ices have had less time to bring the
HDO/H$_2$O ratio back down to the equilibrium ratio at $\geq$100~K.

\subsection{Interferometric continuum images}
\label{s:bure_cont}

\begin{figure}[tb]
\includegraphics[width=8cm,angle=0]{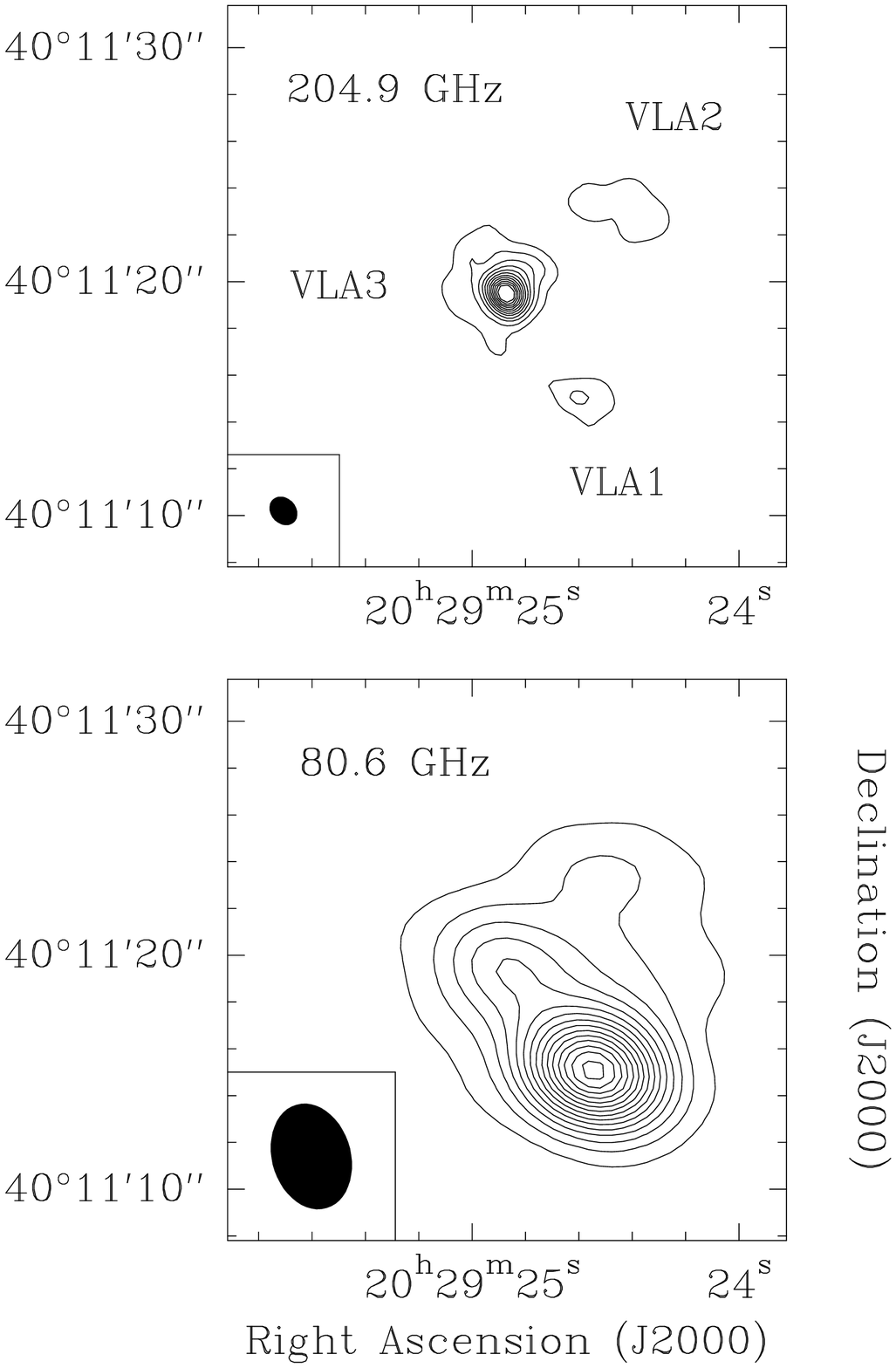}  
  \caption{Maps of the continuum emission of AFGL 2591 made with the IRAM
    interferometer. Contours are drawn every 8~mJy/beam at 205~GHz (top) and
    every 3~mJy/beam at 81~GHz (bottom).  \cut{The strong 81~GHz source is VLA1; the
    strong 205~GHz source is VLA3} \new{Source nomenclature and synthesized
    beam size are indicated}. }
  \label{fig:cont_maps}
\end{figure}

Figure~\ref{fig:cont_maps} presents continuum maps of AFGL 2591, made by
gridding and Fourier transforming the IRAM interferometer data and
deconvolving with the Clean algorithm.  Using uniform weight, the size
(FWHM) of the synthesized beam is (1.33$\times$1.07)$''$ at PA=47$^\circ$
at 205~GHz and (4.60$\times$3.38)$''$ at PA=74$^\circ$ at 81~GHz.  The rms
noise level of the maps is 0.34~mJy/beam at 81~GHz and 1.3~mJy/beam at
205~GHz.

\begin{table*}[tb]
\caption{Positions, deconvolved sizes, and strengths of continuum sources
  in the AFGL 2591 region, detected with the interferometer. Numbers in brackets denote
  uncertainties in units of the last decimal.} 
\label{t:bure_c}
\begin{tabular}{lcccccc}

\hline
\hline

Component$^a$  & R.A. (J2000) & Dec. (J2000) & Major axis & Minor axis & Pos.~angle & Flux density \\
               & hh mm ss     & dd mm ss     & arcsec     & arcsec     & deg.       & Jy \\

\hline

\multicolumn{7}{c}{\it 81 GHz} \\
\\
VLA1 & 20:29:24.5507(10) & 40:11:15.250(10) & 2.64(4) & 2.36(6) & --22(8) & 0.061(1) \\
VLA2 & 20:29:24.4795(50) & 40:11:22.460(46) & ... & ... & ... & 0.009(1) \\
VLA3 & 20:29:24.8917(30) & 40:11:19.687(28) & ... & ... & ... & 0.016(2) \\
\\
\multicolumn{7}{c}{\it 205 GHz}\\
\\
VLA1 & 20:29:24.5721(28) & 40:11:14.669(29) & 2.13( 9) & 1.29( 9) & 90(4) & 0.061(2) \\
VLA2 & 20:29:24.4394(49) & 40:11:23.505(51) & 3.24(17) & 2.21(16) & 90(5) & 0.065(4) \\
VLA3 & 20:29:24.8694( 3) & 40:11:19.498( 5) & 1.08( 2) & 0.85( 1) & 00(3) & 0.194(1) \\

\hline

\end{tabular}

\medskip

{\scriptsize a}: Nomenclature from \citet{trinidad:gl2591}.

\end{table*}

The 205~GHz map shows three sources, and Table~\ref{t:bure_c} lists
their properties, derived by fitting two-dimensional Gaussians to the
\textit{u,v} data. The strongest, Eastern source (VLA3) coincides
with the `dust peak' and the infrared source AFGL 2591. The
South-Western source VLA1 is a compact \hii\ region which dominates the
field at frequencies $\ltsim$100~GHz. The weakest, North-Western source
coincides with feature `VLA2' in low-frequency ($<$10~GHz) VLA maps.
Our 81~GHz map (Fig.~\ref{fig:cont_maps}, bottom) shows the same three
features, but their emission is blended due to the lower angular
resolution.  These results are consistent with previous mapping at
similar frequencies with the OVRO interferometer \citep{vdtak:gl2591}.
However, the Bure data have higher sensitivity, making these the first
detections of the NW and SW \hii\ regions at frequencies $>$200~GHz.
On the other hand, the OVRO data have higher resolution, so that the
infrared source and the \hii\ regions are well separated at 87~GHz.
Source VLA2 is visible in the OVRO 87~GHz data, but not firmly detected.

To study the physical nature of the continuum sources, 
we calculate their millimeter-wave spectral indices $\gamma$, defined as
$S_\nu \propto \nu^\gamma$. To do so we combine the 87~GHz data from OVRO with the
205~GHz data from IRAM. 
The \textit{uv} coverage of these telescopes is similar,
and the effect of `missing flux' on their comparison is probably small.
The result is $\gamma$=0.0$\pm$0.05 for VLA1,
$\gamma$=$+$2.1$\pm$0.2 for VLA2 and $\gamma$=$+$2.7$\pm$0.1 for VLA3.  These
values indicate optically thin free-free emission for VLA1, optically thick dust
or free-free emission in VLA2, and optically thin dust emission in VLA3.  
The brightness temperatures of the 205~GHz sources are 0.7~K for VLA1, 0.3~K for
VLA2 and 6.1~K for VLA3. These values are much lower than the expected physical
temperatures of either ionized gas or dust, and indicate either a low optical
depth or a small filling factor.

For the `dust' source VLA3, we calculate the mass from the observed 205~GHz flux
density, assuming a dust temperature of 100~K, \new{a standard gas to dust ratio
of 100 } and a mass opacity of 0.9~cm$^2$
per gram of dust (\citealt{ossenkopf:opacities}; \citealt{henning:massive}).
The result is 0.8~\msol, which is inversely proportional to the assumed temperature.
Interestingly, the spectral index of VLA3 indicates a value of the dust opacity
index of $\beta \approx 1$, which is smaller than the `usual' value of $\approx$2
and suggests grain growth. This process is thought to occur in circumstellar
disks, which is not inconsistent with the observed elongated shape of VLA3.
The very compact 43~GHz emission from ionized gas seen by \citet{vdtak:qband} 
may then originate in the ionized inner part of the disk, a disk atmosphere,
or a disk wind (\citealt{hollenbach:photevap}; \citealt{lugo:photevap}).

\subsection{Interferometer observations of \hho, HDO and \soo\ line emission}
\label{s:bure_lines}


\begin{figure}[tb]
\includegraphics[width=8cm,angle=0]{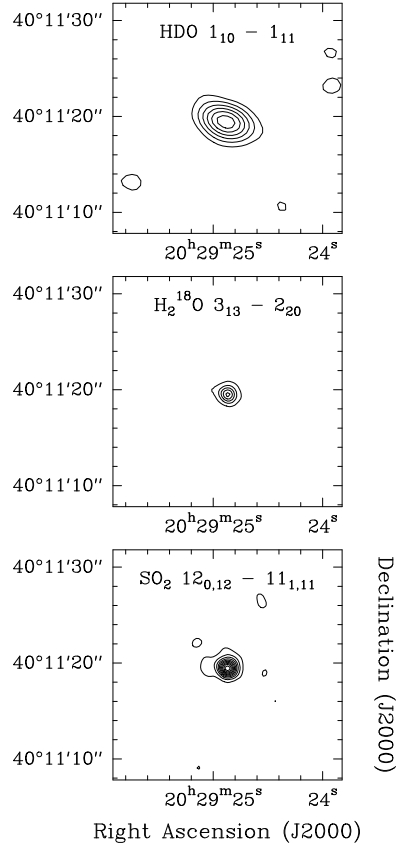}  
  \caption{Interferometric maps of line emission toward AFGL 2591 at the
    central velocity. For HDO (top), first contour and
    contour step are 30~mJy/beam. For \hhoe\ (middle) and
    \soo\ (bottom), first contour is 0.15~Jy/beam and
    contour step is 0.3~Jy/beam.}
  \label{fig:line_maps}
\end{figure}

\begin{figure}[tb]
\includegraphics[width=8cm,angle=0]{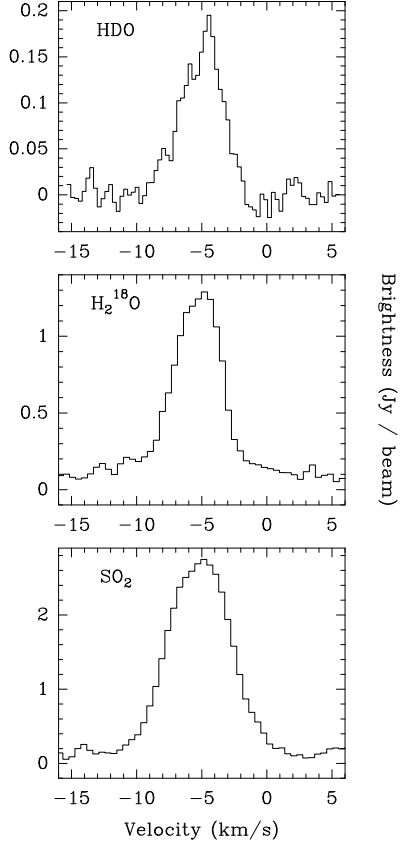}  
  \caption{Interferometric spectra of line emission toward AFGL 2591 taken at
    the image maxima.} 
  \label{fig:bure_spectra}
\end{figure}

Figure~\ref{fig:line_maps} shows the maps of the HDO, \hhoe\ and \soo\
line emission observed with Bure. The beam sizes are the same as those
of the continuum maps at that frequency. The rms noise levels of the line
maps are 9~mJy/beam for HDO, 23~mJy/beam for \hhoe, and 50~mJy/beam for \soo.
In the case of \soo, the noise is limited by dynamic range problems.
%
The line maps show compact emission, coincident with the `dust peak' of AFGL
2591. Columns 2--6 of Table~\ref{t:bure_l} list the position, size and shape of
the emission, obtained by fitting 2D Gaussians to the \textit{u,v} data.  
%
%
Figure~\ref{fig:bure_spectra} shows spectra of the line emission,
taken at the peak positions of the images. 
Columns 7--9 of Table~\ref{t:bure_l} list the central
positions, widths, and peak strengths of the lines, obtained by fitting Gaussian
profiles to the spectra at the image maxima.
The central velocities of the lines are consistent with
the values measured at the 30-m telescope (Table~\ref{t:width}). 
The width of the \soo\ line is consistent with the single-dish value, while the
\hhoe\ line is 23\% broader and the HDO line 50\% broader. 
%
Within the errors, all of the single-dish flux is recovered by the
interferometer: apparently, both telescopes trace the same gas, which has a
compact ($\ltsim$1$''$) distribution.
\new{Given the constraints on the source geometry from the comparison
  of single-dish and infrared column density estimates (\S~\ref{s:cold}), 
  we conclude that the \hhoe\ and HDO emitting regions have sizes of 
  $\sim$1$''$.}

\begin{table*}[tb]
\caption{Positions, deconvolved sizes, velocities, and strengths of emission lines
  detected toward AFGL 2591 with the interferometer. Numbers in brackets
  denote uncertainties in units of the last decimal. } 
\label{t:bure_l}
\begin{tabular}{lllllrrrr}

\hline
\hline

Line  & R.A. (J2000) & Dec. (J2000)  & Major axis & Minor axis & Pos.~angle & \vlsr\ & \dv\ & Peak $T_B$ \\
      & hh mm ss     & dd mm ss      &   arcsec   & arcsec     & deg        & \kms   & \kms & K \\

\hline

HDO     & 20:29:24.872(11) & 40:11:19.48(10) & $\ltsim 1$ & $\ltsim 1$ & ...& --4.92(5) & 3.63(10) &  2.05( 5) \\
\hhoe\  & 20:29:24.8691(6) & 40:11:19.493(8) & 0.81(2) & 0.71(3) & --47(12) & --5.29(3) & 4.12( 7) & 24.26(36) \\
\soo\   & 20:29:24.8688(4) & 40:11:19.363(6) & 1.00(2) & 0.91(2) & --28( 9) & --5.16(2) & 5.44( 5) & 53.93(40) \\

\hline

\end{tabular}

\end{table*}

\subsection{Velocity structure of the line emission}
\label{s:velo}

\begin{figure}[tb]
\includegraphics[height=12cm,angle=0]{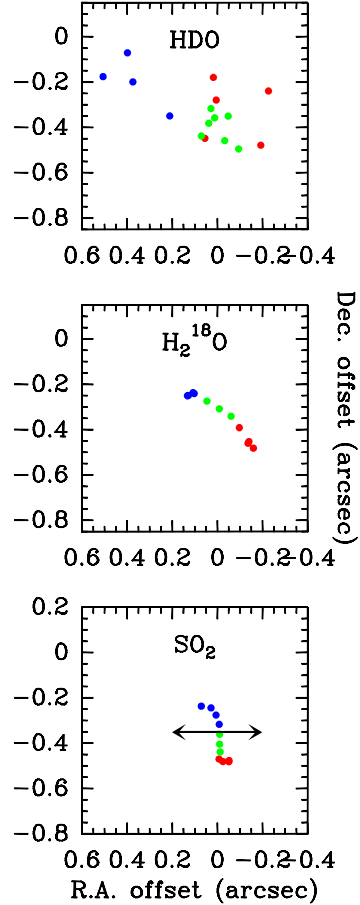}  
  \caption{Position of the emission peak in AFGL 2591 in each channel, as offset from the
    phase center of the Bure observations, for HDO (top), \hhoe\ (middle) and
    \soo\ (bottom). Colour coding corresponds to velocity offset: red =
    redshifted (--\vlsr=2.5 -- 4~\kms), green = line center (--\vlsr=4.5 --
    6~\kms), blue = blueshifted (--\vlsr=6.5 -- 8~\kms). \new{The arrow in the
    bottom panel indicates the orientation of the large-scale CO
    outflow.} }
  \label{f:velo}
\end{figure}

Channel maps of \hhoe, HDO and \soo\ do not show clear changes of the emitting
structure with velocity, but the signal
to noise ratio of the Bure line data is high enough to locate the emission peak to a
fraction of a synthesized beam width. Therefore, to study the velocity structure
of the compact molecular gas, we have determined the position of the emission
peak in each channel by a fit to the \textit{(u,v)} data,
and Fig.~\ref{f:velo} shows the result. The \hhoe\ and \soo\ lines show a clear
velocity gradient, where the redshifted gas is located in the South-West and the
blueshifted gas in the North-East. The HDO shows the same trend, but not as clearly
due to the lower angular resolution. In each case, the gradient is smooth,
which suggests that the emission traces a physically coherent structure.
Since the central position of the line emission coincides with that of the dust
continuum peak, and its central velocity with that of the large-scale molecular
envelope, the most plausible origins of the velocity gradient are outflowing motions
in a bipolar cavity, or rotation in a circumstellar disk.

\new{One central prediction of disk accretion models of low-mass star
  formation is that the outflow axis is perpendicular to the disk
  plane. }
The orientation of the structure seen in Figure~\ref{f:velo} does not agree with
that of the large-scale outflow which is known to emanate from AFGL 2591.
The position angle of the velocity gradient is 39$^\circ$ in \hhoe\ and
67$^\circ$ in \soo\ (measured East from North). The value for HDO is 13$^\circ$,
but this number is uncertain due to the large scatter in the data points, and
not inconsistent with the value for \hhoe.
In contrast, single-dish CO 3--2 and HCO$^+$ 4--3 mapping
\citep{hasegawa:gl2591} shows an outflow of size 90$\times$20$''$, embedded in an
arcminute-scale outflow seen before in lower-$J$ CO lines (e.g, \citealt{lada:gl2591}).
The East-West orientation agrees with the positions of Herbig-Haro
objects and spots of shock-excited \hh\ emission (\citealt{tamura:gl2591};
\citealt{poetzel:h-h}). 
%
On smaller scales, VLBI observations of \hho\ masers by \citet{trinidad:gl2591}
show a shell-like structure which they interpret as an outflow cavity. The spots
are spread over 0.02$''$, in an elongated structure oriented about 20$^\circ$
West from North.

On the other hand, if the line emission seen with Bure is due to a disk, the
inclination is probably close to face-on. For example, the Gaussian fits
presented in Table~\ref{t:bure_l} imply axis ratios of 0.90$\pm$0.02 for \hhoe\
and \soo, corresponding to an inclination of (26$\pm$3)$^\circ$. 
A face-on disk is also consistent with the much higher outflow
velocities seen in CO mid-IR absorption than in mm emission.
The outflow is thus directed almost along the line of sight.

The total magnitude of the velocity gradient is 4.6~\kms\ over an
\cut{angle} \new{offset } of
0.3$''$.  Adopting a distance of 1~kpc and assuming the above inclination angle,
the implied rotation period is $\sim$1000~yr. The central star has a mass of
$\approx$16~\msol, based on the luminosity of the region \citep{vdtak:qband}.
For this orbital period and stellar mass, Kepler's third law implies an orbital
radius of 250~AU, which is within a factor of 2 from the measured value.
This agreement does not prove that the line emission originates in a rotating
disk, but it does indicate that the gas motion is controlled by stellar gravity.
Higher resolution observations are necessary to resolve the velocity field
of the compact line emission.

\section{Radiative transfer analysis}
\label{s:radtrans}

Dividing the single-dish column densities in Table~\ref{t:cold} by the
$N$(\hh) values based on \smm\ observations of dust continuum and
C$^{17}$O lines in $\approx$15$''$ beams \cut{\citep{vdtak:sulphur}}
\new{(Table~\ref{tab:samp}) } leads to beam-averaged abundances of
2$\times$10$^{-10}$ -- 1$\times$10$^{-9}$ for HDO and
2$\times$10$^{-7}$ -- 2$\times$10$^{-6}$ for \hho.
\cut{Our data may thus sample $\sim$100\% of gas of the sort seen with
SWAS, which has an \hho\ abundance of $\sim$10$^{-7}$, or  $\sim$0.1\%
of the gas seen with ISO where \hho\ is a major carrier of oxygen and 
has an abundance of $\sim$10$^{-4}$.}
On the other hand, the interferometer data of AFGL 2591 indicate a
\hho\ column density of \pow{3.4}{19}~\scm, assuming \new{optically
  thin emission with }
\txc=100~K, while the 205~GHz continuum indicates $N$(\hh)=\pow{2.1}{24}~\scm\ 
for a dust temperature of 100~K.
The \hho/\hh\ column density ratio of \pow{1.6}{-5} is $\gtsim$10$\times$ higher
than estimated from the single-dish data.
To resolve this discrepancy, we have run radiative transfer models of the line emission.

\subsection{Model description}
\label{s:hst}

The column density ratios evaluated in the previous section are crude
estimates of the molecular abundances because they do not take excitation
gradients along the line of sight into account. In protostellar envelopes and disks, where
density and temperature vary over orders of magnitude, these gradients are very
significant. Furthermore, column densities and their ratios do not give insight
into the location of the gas along the line of sight.
To estimate more accurate abundances,
the line emission of \hhoe\ and HDO in our sources has been modeled
with the Monte Carlo radiative transfer program of
\citet{hvdt:hst}\footnote{\tt http://www.mpifr-bonn.mpg.de/staff/fvandertak/ratran/}. 
The application of this program to \hho\ has been explicitly tested in
a dedicated benchmark campaign \citep{vdtak:h2o-benchmark}\footnote{\tt
http://www.mpifr-bonn.mpg.de/staff/fvandertak/H2O/}.
Spectroscopic and collisional input for the modeling comes
from the molecular database by \citet{schoeier:moldata}\footnote{\tt
  http://www.strw.leidenuniv.nl/$\sim$moldata/}. 
Besides collisional excitation, dust radiation is taken into account using grain
properties from \citet{ossenkopf:opacities}, Model~5.
Radial profiles of the temperature and density of the envelopes of our
sources were determined by \citet{vdtak:massive} based on single-dish maps
of dust continuum and molecular line emission at \smm\ wavelengths.

\subsection{Constant-abundance models}
\label{s:mres}

\begin{figure}[tb]
\includegraphics[width=6cm,angle=-90]{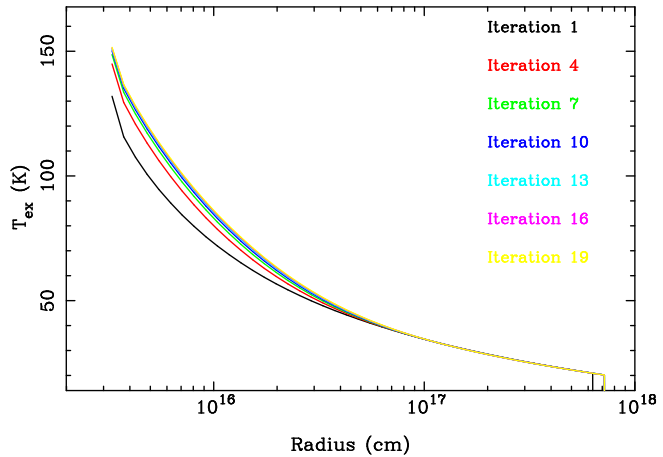}  
  \caption{Excitation temperature of the \hho\ 3$_{13}$--2$_{20}$ line as a function of
    radius for the AFGL 2591 model. The curves for the last few iteration stages
    in the Monte Carlo calculation coincide, which shows that the simulation has
    converged.}
  \label{f:cnv_plot}
\end{figure}

We first assume that \hho\ and HDO are distributed evenly over the protostellar envelopes.
Table~\ref{t:abun} reports the results of these models. 
Stringent convergence criteria had to be applied, as the upper (emitting) level
of the \hhoe\ line has a relative population of only $\sim$10$^{-7}$ in the outer
parts of the envelopes. Nevertheless, Figure~\ref{f:cnv_plot} shows that the
calculations are well converged.
We calculate an optical depth for the \hhoe\ line of $\sim$10$^{-2}$. Maser
action, as often observed in the \hho\ 183~GHz line, does not occur at \hhoe\
abundances of $\sim$10$^{-8}$.
The derived abundances are factors of $\sim$30 higher than the line-of-sight
averages from \S~\ref{s:cold}, due to the strong excitation gradients in these
sources. 

\begin{table}
\caption{Envelope-averaged abundances$^a$ of HDO and \hho\ relative to
    \hh, derived from radiative transfer models.}
\label{t:abun}
\begin{tabular}{lccc}

\hline
\hline

Source        & HDO       & \hho       \\
              & 10$^{-8}$ & 10$^{-5}$  \\

\hline

W3 IRS5       & 1 & 0.8  \\
W33A          & 3 & 1.0  \\
AFGL 2591     & 2 & 6.0  \\
NGC 7538 IRS1 & 3 & 0.8  \\

\hline

\end{tabular}

\medskip

{\scriptsize a}: Average value over envelope

\end{table}

\subsection{Jump models}
\label{s:jres}

Although the constant-abundance models in the previous section fit the strength
of the \hhoe\ line and the average strength of the HDO lines well, they have
several shortcomings. First, they predict FWHM sizes of the \hhoe\ and HDO emission in AFGL
2591 of 4--5$''$, significantly larger than the observed $\ltsim$1$''$. Second,
the fit residuals of HDO are correlated with energy level, in the sense that
low-excitation lines tend to be overproduced and high-excitation lines
underproduced. Both effects suggest that \hho\ and HDO are not distributed
evenly throughout the sources, but have enhanced abundances in the warm inner
envelopes. Single-dish spectra and interferometer maps of H$_2$CO and CH$_3$OH
\citep{vdtak:meth} and of SO and SO$_2$ \citep{vdtak:sulphur} show the same effect.

Assuming that the water is produced by evaporation of icy grain mantles, 
we have run `jump' models for \hhoe\ in AFGL 2591. 
The parameters of this model are the abundance in the warm inner region and the
temperature at which the ice evaporates.
%
Laboratory studies indicate that the evaporation temperature lies in the 90 --
110~K range, depending on ice composition and structure
(\citealt{pontoppidan:co_ice}; \citealt{fraser:co_ice}).
With only one transition of \hhoe\ observed, it is not possible
to constrain both parameters simultaneously, so we
have initially fixed the boundary of the inner region at the 100~K point.
The water abundance is assumed negligible outside this radius, which for AFGL
2591 lies at 2000~AU. The upper level of the observed transition is too high to
set useful limits on the water abundance in the outer envelope.
We find that the observed line flux and source size are reproduced for \hho/\hh\ 
= 1.4$\times$10$^{-4}$ which represents a major fraction of the available
oxygen.

Alternatively, the ice mantles may evaporate at a somewhat different temperature.
We have run a model with the \hho\ abundance in the warm gas fixed at
2$\times$10$^{-4}$, and varied the radius of the inner region to match the
data. The best-fit model of this kind has the ice evaporating at $T$=115~K.
We consider both \hho\ jump models plausible; multi-line observations of
\hhoe\ are needed to rule out either model.
\new{This result is consistent with modeling of the SWAS data of AFGL
  2591 \citep{boonman:models} which indicates an evaporation
  temperature of 90 -- 110~K.} 

\new{The present data do not constrain the abundance of \hho/\hh\ in
  the outer envelope very well. \citet{boonman:models} derive an upper limit
  of $\sim$10$^{-8}$ from a combined analysis of ISO-SWS, ISO-LWS and
  SWAS data which cover a range of energy levels.
  However, \cut{these data suffer from lack of spectral
  resolution, and} such a low \hho\ abundance would imply an
  HDO/\hho\ ratio of $\sim$unity in the outer envelope, which is
  implausibly high for this type of object.
Our best-fit model to the SWAS data has \hho/\hh\ $\sim$10$^{-6}$ in
  the outer envelope, inconsistent with the results by Boonman et al,
  but uncertain because based on only one transition.
  Clearly, \textit{Herschel}-HIFI data are needed to settle this issue.
}

For HDO, `jump' models were run for each source except W3 IRS5, where too few
lines were observed to constrain such models. In these models, the jump occurs
at the fixed location of $T$=100~K, and the HDO abundance inside and outside this
radius are allowed to vary independently. 
The `jump' models reproduce all the single-dish line fluxes to within 50\%,
which is about the expected error margin. 
Table~\ref{t:jump} reports the results of these models.
For AFGL 2591, the size of the 81~GHz line emission measured with Bure acts as
an extra constraint.
The best-fit constant-abundance model predicts an emitting region of
$\approx$5$''$ FWHM, whereas the jump model predicts $\approx$2$''$, consistent
with the measured value, so that this latter model is favoured.



The `jump' models indicate that the HDO/\hho\ ratio is
\cut{$\approx$\pow{1.4}{-3}} \new{$\approx$\pow{5}{-4} } in
the inner region. This value corresponds to an enhancement over the interstellar
D/H ratio by 100, and is consistent with the observational limits on the
HDO/\hho\ ratio in the solid state. We conclude that the bulk of the HDO and
\hho\ seen with the 30m telescope and the Plateau de Bure interferometer is
evaporated ice.

\begin{table*}
\caption{Abundances of HDO and \hho\ in the inner and outer envelopes of the sources,
         derived from radiative transfer models.}
\label{t:jump}
\begin{tabular}{lrrcccc}

\hline
\hline

Source & \multicolumn{2}{c}{HDO/\hh} & \multicolumn{2}{c}{\hho/\hh} & \multicolumn{2}{c}{HDO/\hho} \\
       & \multicolumn{2}{c}{10$^{-9}$}    & \multicolumn{2}{c}{10$^{-4}$}     & \multicolumn{2}{c}{10$^{-4}$} \\
       & inner & outer          & inner & outer          & inner & outer   \\
\hline

W33A          & 200 & 10        & ...           &  ...              &  ... &  ... \\
AFGL 2591     & 100 & 4         & 1.4 -- 2 $^a$ & 10$^{-2}$ -- 10$^{-4}$ $^b$    & 5    & 40 -- 4000 \\
NGC 7538 IRS1 & 100 & 20        & ...           & ...    &  ... &  ... \\

\hline

\end{tabular}

\medskip

{\scriptsize a}: Values for ice evaporating at $T$=100~K and $T$=115~K.

{\scriptsize b}: From \citet{boonman:models}.

\end{table*}

\section{Discussion}
\label{s:disc}

\subsection{Geometry of AFGL 2591: A massive circumstellar disk?}
\label{ss:geom}

An important result from this study is that the \hhoe\ 203~GHz emission from
AFGL 2591 is very compact (800~AU diameter) and coincident within the errors
with the continuum emission from VLA3. \cut{From the 1.3~mm continuum data, we
estimate a mass of 0.8~\msol, assuming a dust temperature $T_d$ of 100~K and a
mass opacity of the dust of 0.9~cm$^2$ per gram of dust
\citep{ossenkopf:opacities}. For $T_d$=50 and 200~K, the mass estimates are 1.8
and 0.4~\msol. These masses are of order}
\new{The 1.3~mm continuum data indicate a mass of 0.8~\msol, which is
  $\approx$}5\% of the stellar mass of 
$\approx$16~\msol\ \citep{vdtak:qband}. One possible interpretation of these
data is that we are observing the circumstellar disk surrounding the central
star of AFGL 2591. The mass ratio seems plausible for a young protostar and
given the observed axis ratios of the line and continuum emission
(Tables~\ref{t:bure_c} and~\ref{t:bure_l}), this putative
disk is likely to be close to face-on (inclination 26 -- 38 degrees).
This inclination is compatible with the spectral energy distribution of the
source in the near- to mid-infrared range (e.g., \citealt{whitney:geometry}).
However, there are certainly other possible interpretations of the data, at
least as far as the line emission is concerned.

Another significant result from the interferometer observations is that neither
\hhoe\ nor \soo\ is distributed in a spherically symmetric fashion around the
central star. Figure~\ref{f:velo} shows that the red and blue wings of the lines
are offset from one another along a roughly NE -- SW direction. The orientation
seems to be somewhat different for water and \soo. While one clearly needs much
better angular resolution to interpret the data, we conclude that the velocity
gradient in Fig.~\ref{f:velo} is probably due to either disk rotation or the
effect of the interaction of an outflow with the sides of a cavity (presumably
directed toward us). \soo\ is a known outflow indicator \citep{schilke:345survey}
and, more in general, is found associated with shocked gas. It seems likely that
in a wide-angle wind inclined at a moderate angle to the line of sight,
different outflow tracers will show different orientations in a representation
such as that of Fig.~\ref{f:velo}. Such differences in orientation may
particularly occur if the jet driving the outflow were to be precessing, which
VLBI observations of \hho\ 22~GHz indicate for AFGL 2591 \citep{trinidad:gl2591}.
Although higher resolution imaging is required to settle this issue, we note
that the column densities of \hho, \soo\ and \hh\ are well above the values in
other massive molecular outflows \citep{beuther:outflow}, making this scenario
implausible.

Near-infrared speckle imaging of AFGL 2591 by \citet{preibisch:afgl2591}
supports our picture of the geometry of the source. These images (as well as
older ones) show several loops of emission extending due West from the central
source, with major axes of $\approx$10$''$ and axis ratios of $\approx$3. This
emission presumably traces a limb-brightened cavity around the approaching
outflow lobe. Since the outflow is expected to be perpendicular to the disk, the
Western orientation of the loops is nicely consistent with the N--S orientation
of the 205~GHz continuum emission. The near-infrared images thus indicate that
the Western part of the disk is tilted away from us. The axis ratios of the
loops are larger than that of the disk in the Bure images, as expected because
the outflow is an intrinsically elongated structure, unlike the disk.
The proposed outflow orientation is also consistent with the other observations
mentioned in \S~\ref{s:velo}.
In these data, the general direction of the outflow is East-West, with some
diversity among tracers as expected in the proposed pole-on orientation of the
system.

It also is worth noting that the source-averaged $N$(\hho) toward AFGL 2591,
corrected to a 0$''$.8 source size, is 5$\times$10$^{19}$~\scm, two orders of
magnitude higher than the value derived from ISO 6~\mic\ data \citep{boonman:h2o}.
This difference again suggests an asymmetrical distribution of material around
the central source as, for example, in a disk geometry. Along the line of sight
to the central source, one observes across the outflow cavity but column
densities are much larger in the disk or perhaps (see above) toward the cavity
edges where outflow and inner envelope interact. In this scenario, the high excitation
temperatures derived from the ISO data indicate that the cavity is hot (several 100~K).

\begin{figure}[tb]
\includegraphics[height=9cm,angle=-90]{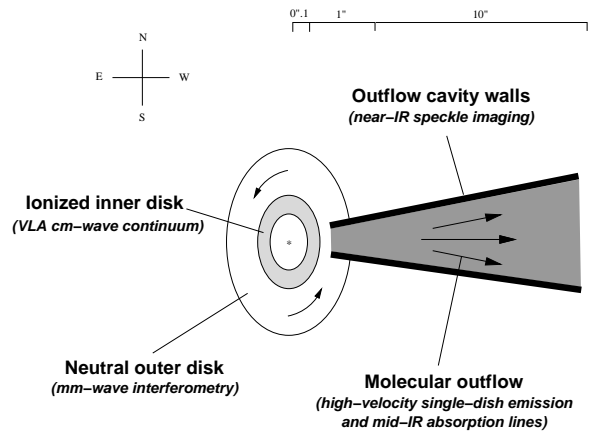}  
  \caption{Sketch of the inner part of the AFGL 2591 region, as
    projected on the sky, with the observational characteristics of
    each physical component indicated. At a distance of 1~kpc, 1$''$
    corresponds to 1000~AU.}
  \label{f:view}
\end{figure}

\new{Figure \ref{f:view} summarizes our combined interpretation of the
new and the existing observations of the central region of AFGL
2591. The circumstellar disk and the molecular outflow are embedded in
a large-scale molecular envelope, observed as the low-velocity
single-dish emission and mid-infrared absorption. For clarity, this
large-scale envelope is not drawn here, but it is depicted in Fig.~11
of \cite{vdtak:gl2591}, along with additional large-scale features. 
Note that part of the cm-wave emission observed with the VLA may arise
in the base of the outflow.}

\subsection{Chemistry of water}
\label{ss:chem}

It is also of great interest that the `jump' models in Section~\ref{s:jres} show
that our observations are consistent with the idea that the \hhoe\ emission
traces gas with a temperature above 100~K. Moreover, the \hho\ abundances derived
from these models are \cut{(to within a factor of 2)} equal to \new{or
  a few times higher than } those estimated for \hho\ 
ice in these sources (\citealt{vdishoeck:faraday}; \citealt{boonman:models}). 
Thus, we believe that the 3$_{13}$--2$_{20}$ transition of \hhoe\ can
be used to trace the behaviour of high temperature gas where water ice
has evaporated. Since the interferometer recovers all single-dish
\hhoe\ line flux, we conclude that the \hhoe\ emission appears to be
an excellent tracer of the inner $\sim$1000~AU of protostars.
In our model of AFGL 2591, the mass inside the 100~K point is 0.2~\msol,
similar to the mass derived from the Bure continuum data.

The warm-up of the central region to 100~K is evidently sufficiently recent that
gas-phase chemistry has not had time to modify the abundances substantially.
The \hho/\hhhop\ ratio in W3~IRS5 is close to the equilibrium value of 1000
\citep{phillips:h3o+}. For W33A, AFGL 2136, AFGL 2591, S140 IRS1 and NGC 7538
IRS9, our unpublished JCMT observations of the \hhhop\ 364~GHz line
indicate upper limits of \tmb$<$0.13 -- 0.21~K. Because of the large Einstein A
coefficient of the transition, its excitation temperature is probably
significantly below that of \hho\ and HDO. Assuming \txc=25~K and an ortho/para
ratio of 2, our upper limits on $N$(\hhhop) are (4--7)$\times$10$^{13}$~\scm. These
numbers correspond to lower limits on the \hho/\hhhop\ ratio of 2000 in W33A and
6000 in AFGL 2591. 
Thus, in these sources, gas-phase chemistry seems not to have had time to return
the \hho/\hhhop\ ratio to its equilibrium value since the evaporation of the
grain mantles.

\def\tion{$t_{\rm ion}$}

The time scale \tion\ to reach chemical equilibrium between \hho\ and \hhhop\ can be
estimated by realizing that \hhhop\ is produced in reactions between \hho\ and
molecular ions, in particular HCO$^+$, He$^+$ and \hhhp. Destruction of \hhhop\ is by
dissociative recombination, which for 25\% re-forms water, but for 60\% makes OH
and 15\% O \citep{jensen:h3o+}.
We thus estimate \tion\ as
$(\alpha_L n_X)^{-1}$, where $\alpha_L$ is the Langevin reaction rate of
$\sim$10$^{-9}$ \ccms, and the concentration of molecular ions $n_X$ is given by
the balance of cosmic-ray ionization on the one hand and
reactions with CO and O on the other. Models by \citet{vdtak:zeta} indicate $n_X \sim
10^{-4}$~\ccm, so that \tion$\sim$\pow{3}{5}~yr.  We conclude that the evaporation of
grain mantles in AFGL 2591 has taken place less than $\sim$0.1~Myr ago.
The time scale may be even shorter, given the mass loss rate of
$\sim$10$^{-4}$~\msol\,yr$^{-1}$ measured in the CO outflow
\citep{hasegawa:gl2591}. Assuming that the 0.8~\msol\ disk accretes at the same
rate, the disk lifetime is only $\sim$10$^4$~yr. \new{This value
  is similar to the age estimate of \pow{3}{4}~yr from multi-species
  chemical modeling of the envelope of AFGL 2591 \citep{doty:massive}.}

\begin{figure}[tb]
\includegraphics[height=9cm,angle=0]{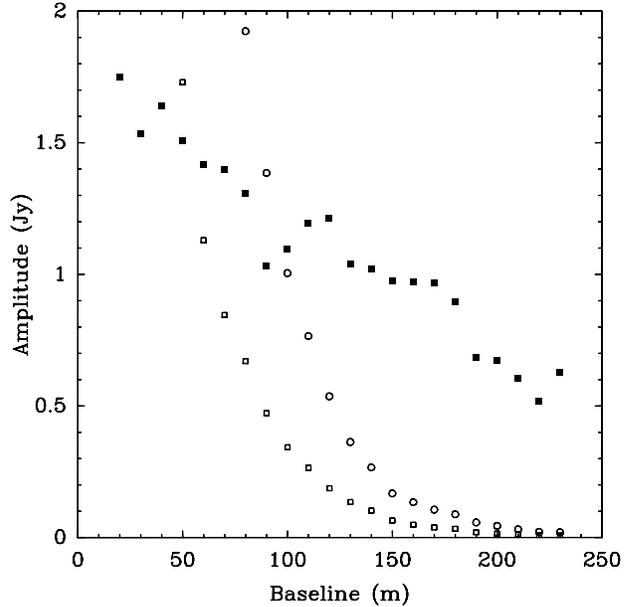}  
  \caption{Filled squares: Visibility amplitude of \hhoe\ line
    emission observed toward AFGL 2591 with the Plateau de Bure
    interferometer, integrated over velocity and binned. Superposed
    are model points for constant abundance (open squares) and for the
    `jump' model (open circles).}
  \label{f:uvmodel}
\end{figure}

It may be asked whether the Monte Carlo treatment of radiative transfer in
\S~\ref{s:radtrans} is useful given our uncertainty on the geometry. In fact, we
think that our spherically symmetric model is adequate to determine the mass of
warm ($>$100~K) gas necessary to explain the observations, and this mass is, for
an essentially optically thin line, geometry independent. However, understanding
line profiles as well as observations at still higher angular resolution will
require a more sophisticated treatment such as axisymmetric modeling.

\new{
Figure~\ref{f:uvmodel} illustrates the limitations of our models by
comparing them to the interferometer data in the \textit{uv} plane.
Although both models reproduce the total flux and the jump model 
also the source size as estimated through Gaussian fits,
neither model reproduces the observations in detail.
This discrepancy hints at the existence of additional geometrical 
structure which is not present in the model.
The two models predict different source sizes, but
the same overall emission shape, probably because
the \hhoe\ line is a tracer of warm gas.
}

Finally, our HDO data and the implied HDO/\hho\ abundance ratios are interesting
in combination with the result from the `jump' models that the gas-phase water
abundance is consistent with material which has recently evaporated off grains.
Thus the observed HDO/\hho\ should reflect the ratio of these species in the
solid state and indeed, our result is consistent with current limits on HDO ice.

\subsection{Comparison with low-mass protostars}
\label{ss:lmpo}

The observations of H$_2^{18}$O presented here have shown
that in the studied high-mass protostars the water emission likely
originates in the envelopes, and that the H$_2$O abundance jumps in
the inner warm region where the grain mantles sublimate. A similar
analysis of the ISO-LWS spectra of two low-mass protostars 
(\citealt{ceccarelli:16293}; \citealt{maret:h2o}) shows that
also in those cases, a jump in the water abundance at approximatively the radius
where the dust temperature reaches 100~K is needed to explain the observed
far-infrared water line spectrum. However, the water abundance in the warm gas
is strikingly different: $\sim$10$^{-4}$ in the high-mass, and
$\sim$3$\times$10$^{-6}$ in the low-mass protostars respectively.
Mid-infrared spectra of low-mass protostars give solid water abundances of
$\sim$10$^{-4}$ as for high-mass protostars \citep{pontoppidan:ice_map}.
\cut{2004 (Mapping protostellar ice in Serpens)} It is not obvious why the
water abundance in the warm gas should be only a few percent of the ice abundance.
As \S~\ref{s:jres} discusses, the exact ice evaporation
temperature depends somewhat on the ice composition and structure, but not
enough to make a factor of 100 difference.
Unless the water abundances in the studied low-mass protostellar envelopes are
affected by the large (80$''$) ISO-LWS beam, it seems that the break-down of
evaporated water ice is faster around low-mass than around high-mass
stars. This conclusion is somewhat surprising since water is
destroyed by ion-molecule chemistry and the few available data suggest
that if anything, the ionization rate is higher around high-mass than
around low-mass stars \citep{vdtak:catania}. In the future, \textit{Herschel}-HIFI
data will be helpful to settle this issue. 

HDO emission, practically at all the same frequencies of the present
work, has been observed towards the low-mass protostar IRAS 16293-2422
\citep{parise:hdo}. Similarly to the present work, the authors
analyzed the observations with an abundance jump model. They found
that the abundance of HDO has a jump from a value $\leq 10^{-9}$ in
the outer envelope, where the dust temperature is lower than 100~K, to
$\sim 10^{-7}$ in the inner envelope.  The abundance found in the
region of sublimated ices is therefore very similar in high- and 
low-mass protostars.  However, the HDO/H$_2$O ratio is substantially
different. In high-mass protostars it is at most $3\times 10^{-3}$,
whereas in the low-mass protostar IRAS 16293-2422 it is $\sim 0.03$,
namely ten times larger and close to to the observational limit on 
solid HDO \citep{parise:solid}. This difference is as expected from 
recent measurements of extreme molecular deuteration around low-mass
protostars, where doubly and triply deuterated molecules have been
detected with fractionations of a few percent (for references, see \citealt{vdtak:catania}).
%
%
%
The degree of deuteration in high-mass protostellar envelopes is much lower,
starting from the failure to detect
H$_2$D$^+$ in massive protostars (\citealt{pagani:h2d+}; \citealt{stark:h2d+}). 
Possibly, for high-mass protostars, the very cold
and dense pre-collapse phase where the CO freezes out onto the grain mantles
lasts only a short time.  The present measurement of HDO/H$_2$O in high mass
protostars, compared with the value found in IRAS 16293-2422, confirms this
hypothesis: high-mass protostars show indeed a lower degree of water deuteration.


The water ice on grain surfaces is laid down in relatively low density molecular
cloud gas (10$^3$ -- 10$^4$~\ccm) where H~atoms are as abundant as O~atoms and thus
there is a relatively high probability of \hho\ forming subsequent to O sticking
to a grain.  After formation, the ice presumably stays frozen until the dust is
heated up by protostellar radiation. It is the temperature of this `primordial'
low-density molecular cloud gas that counts for the chemistry of HDO/\hho. The
present observations suggest that this temperature is higher in the Giant
Molecular Clouds producing high-mass stars than in low-mass star-forming regions
such as Taurus and Ophiuchus.

\section{Conclusion and Outlook}
\label{s:conc}

The chemical composition of star-forming matter depends on both temperature and
time. For over a decade, people have tried to use the time dependence to
estimate chemical ages of star-forming regions. These estimates, the so-called
`chemical clocks' remain unreliable, probably due to uncertainties in the
initial conditions as well as in the physical structure (cf.\ \citealt{vdtak:catania}).
This paper has shown that the temperature dependence may instead be used to apply
`chemical filters'. In particular, we have used the \hho\ molecule to image the
material at $T>100$~K, filtering out the surrounding cooler material. The
success of this filter lies in the evaporation of icy grain mantles which
enhance the \hho\ gas-phase abundance by two orders of magnitude at $T>100$~K.

By using this chemical filter, we have shown that the dust and water inside
2000~AU from the central star in the AFGL 2591 region is asymmetrically
distributed.  Most likely, the observations trace a circumstellar disk of
diameter 800~AU which rotates at about Keplerian speed. 
The result of the \soo\ observations is qualitatively the same as for \hhoe, but
differs in the details. In this case, the `chemical filter' may be a bit leaky
because the \soo\ abundance in the large-scale envelope is non-negligible, and
because \soo\ is also abundant in the bipolar outflow of AFGL 2591.
Furthermore, comparison of our results with those for the prototypical low-mass
system IRAS 16293 shows that the \hho\ abundance in the warm gas around young
high-mass stars is much higher, but the HDO/\hho\ ratio much lower than around
low-mass protostars. 

In the future, observations with the PACS camera and the HIFI spectrometer
onboard the \textit{Herschel} space observatory will further refine this picture
\citep{walmsley:paris}. Observations of multiple \hho\ and \hhoe\ lines will
constrain the excitation and ortho/para ratio of water much better than is
possible from ground-based data. The high sensitivity will allow us to measure
the water abundance in each chemical zone (disk/outflow/envelope), not only
around high-mass protostars, but also around lower-mass objects.
\new{Given our estimated outflow contribution of 10 -- 20\% to the \hhoe\ and
HDO lines in HIFI-sized beams, the contribution from outflows to
H$_2^{16}$O spectra from HIFI will probably be much larger, which
should be considered in the planning of the HIFI observations. }
New, deep searches for \hhhop\ in high-mass protostellar envelopes with APEX,
HIFI and ALMA would also be useful. 
Observations of the \hhoe\ and \soo\ lines in AFGL 2591 on longer baselines are
necessary to resolve the velocity field and test other possibilities such as a
binary system, as seen in the W3~(\hho) source by \citet{wyrowski:w3(h2o)}.

\begin{acknowledgement}
%
  We thank the staffs of the IRAM 30m, JCMT, and Plateau de
  Bure telescopes for assisting with the observations,
  especially Jan Martin Winters and J\'er\^ome Pety at IRAM Grenoble.
  The JCMT data were taken in collaboration with Ewine van
  Dishoeck and Annemieke Boonman at Leiden Observatory.
  Holger M\"uller at the University of Cologne kindly provided 
  \hho\ term values to spectroscopic accuracy.

\end{acknowledgement}

\bibliographystyle{aa}
\bibliography{h2o,cc}

\begin{thebibliography}{62}
\expandafter\ifx\csname natexlab\endcsname\relax\def\natexlab#1{#1}\fi

\bibitem[{{Bachiller} {et~al.}(2001){Bachiller}, {P{\' e}rez Guti{\' e}rrez},
  {Kumar}, \& {Tafalla}}]{bachiller:l1157}
{Bachiller}, R., {P{\' e}rez Guti{\' e}rrez}, M., {Kumar}, M.~S.~N., \&
  {Tafalla}, M. 2001, \aap, 372, 899

\bibitem[{{Beuther} {et~al.}(2002){Beuther}, {Schilke}, {Sridharan}, {Menten},
  {Walmsley}, \& {Wyrowski}}]{beuther:outflow}
{Beuther}, H., {Schilke}, P., {Sridharan}, T.~K., {et~al.} 2002, \aap, 383, 892

\bibitem[{{Blake} {et~al.}(1987){Blake}, {Sutton}, {Masson}, \&
  {Phillips}}]{blake:orion}
{Blake}, G.~A., {Sutton}, E.~C., {Masson}, C.~R., \& {Phillips}, T.~G. 1987,
  \apj, 315, 621

\bibitem[{{Boonman} {et~al.}(2003){Boonman}, {Doty}, {van Dishoeck}, {Bergin},
  {Melnick}, {Wright}, \& {Stark}}]{boonman:models}
{Boonman}, A.~M.~S., {Doty}, S.~D., {van Dishoeck}, E.~F., {et~al.} 2003, \aap,
  406, 937

\bibitem[{{Boonman} {et~al.}(2001){Boonman}, {Stark}, {van der Tak}, {van
  Dishoeck}, {van der Wal}, {Sch{\" a}fer}, {de Lange}, \&
  {Laauwen}}]{boonman:hcn}
{Boonman}, A.~M.~S., {Stark}, R., {van der Tak}, F.~F.~S., {et~al.} 2001,
  \apjl, 553, L63

\bibitem[{{Boonman} \& {van Dishoeck}(2003)}]{boonman:h2o}
{Boonman}, A.~M.~S. \& {van Dishoeck}, E.~F. 2003, \aap, 403, 1003

\bibitem[{{Ceccarelli} {et~al.}(2000){Ceccarelli}, {Castets}, {Caux},
  {Hollenbach}, {Loinard}, {Molinari}, \& {Tielens}}]{ceccarelli:16293}
{Ceccarelli}, C., {Castets}, A., {Caux}, E., {et~al.} 2000, \aap, 355, 1129

\bibitem[{{Cernicharo} {et~al.}(1990){Cernicharo}, {Thum}, {Hein}, {John},
  {Garcia}, \& {Mattioco}}]{cernicharo:h2o}
{Cernicharo}, J., {Thum}, C., {Hein}, H., {et~al.} 1990, \aap, 231, L15

\bibitem[{{Dartois} {et~al.}(2003){Dartois}, {Thi}, {Geballe}, {Deboffle},
  {d'Hendecourt}, \& {van Dishoeck}}]{dartois:hdo}
{Dartois}, E., {Thi}, W.-F., {Geballe}, T.~R., {et~al.} 2003, \aap, 399, 1009

\bibitem[{{Doty} {et~al.}(2002){Doty}, {van Dishoeck}, {van der Tak}, \&
  {Boonman}}]{doty:massive}
{Doty}, S.~D., {van Dishoeck}, E.~F., {van der Tak}, F.~F.~S., \& {Boonman},
  A.~M.~S. 2002, \aap, 389, 446

\bibitem[{{Fraser} {et~al.}(2005){Fraser}, {Bisschop}, {Pontoppidan},
  {Tielens}, \& {van Dishoeck}}]{fraser:co_ice}
{Fraser}, H.~J., {Bisschop}, S.~E., {Pontoppidan}, K.~M., {Tielens},
  A.~G.~G.~M., \& {van Dishoeck}, E.~F. 2005, \mnras, 356, 1283

\bibitem[{{Gensheimer} {et~al.}(1996){Gensheimer}, {Mauersberger}, \&
  {Wilson}}]{gensheimer:h2o}
{Gensheimer}, P.~D., {Mauersberger}, R., \& {Wilson}, T.~L. 1996, \aap, 314,
  281

\bibitem[{{Hasegawa} \& {Mitchell}(1995{\natexlab{a}})}]{hasegawa:outflow}
{Hasegawa}, T.~I. \& {Mitchell}, G.~F. 1995{\natexlab{a}}, \apj, 441, 665

\bibitem[{{Hasegawa} \& {Mitchell}(1995{\natexlab{b}})}]{hasegawa:gl2591}
---. 1995{\natexlab{b}}, \apj, 451, 225

\bibitem[{{Helmich} {et~al.}(1994){Helmich}, {Jansen}, {de Graauw},
  {Groesbeck}, \& {van Dishoeck}}]{helmich:w3}
{Helmich}, F.~P., {Jansen}, D.~J., {de Graauw}, T., {Groesbeck}, T.~D., \& {van
  Dishoeck}, E.~F. 1994, \aap, 283, 626

\bibitem[{{Helmich} {et~al.}(1996){Helmich}, {van Dishoeck}, \&
  {Jansen}}]{helmich:hdo}
{Helmich}, F.~P., {van Dishoeck}, E.~F., \& {Jansen}, D.~J. 1996, \aap, 313,
  657

\bibitem[{{Henning} {et~al.}(2000){Henning}, {Schreyer}, {Launhardt}, \&
  {Burkert}}]{henning:massive}
{Henning}, T., {Schreyer}, K., {Launhardt}, R., \& {Burkert}, A. 2000, \aap,
  353, 211

\bibitem[{{Hogerheijde} \& {van der Tak}(2000)}]{hvdt:hst}
{Hogerheijde}, M.~R. \& {van der Tak}, F.~F.~S. 2000, \aap, 362, 697

\bibitem[{{Hollenbach} {et~al.}(1994){Hollenbach}, {Johnstone}, {Lizano}, \&
  {Shu}}]{hollenbach:photevap}
{Hollenbach}, D., {Johnstone}, D., {Lizano}, S., \& {Shu}, F. 1994, \apj, 428,
  654

\bibitem[{{Jacq} {et~al.}(1988){Jacq}, {Henkel}, {Walmsley}, {Jewell}, \&
  {Baudry}}]{jacq:h2o18}
{Jacq}, T., {Henkel}, C., {Walmsley}, C.~M., {Jewell}, P.~R., \& {Baudry}, A.
  1988, \aap, 199, L5

\bibitem[{{Jacq} {et~al.}(1990){Jacq}, {Walmsley}, {Henkel}, {Baudry},
  {Mauersberger}, \& {Jewell}}]{jacq:hdo}
{Jacq}, T., {Walmsley}, C.~M., {Henkel}, C., {et~al.} 1990, \aap, 228, 447

\bibitem[{{Jensen} {et~al.}(2000){Jensen}, {Bilodeau}, {Safvan}, {Seiersen},
  {Andersen}, {Pedersen}, \& {Heber}}]{jensen:h3o+}
{Jensen}, M.~J., {Bilodeau}, R.~C., {Safvan}, C.~P., {et~al.} 2000, \apj, 543,
  764

\bibitem[{{J{\o}rgensen} {et~al.}(2005){J{\o}rgensen}, {Sch{\" o}ier}, \& {van
  Dishoeck}}]{jorgensen:methanol}
{J{\o}rgensen}, J.~K., {Sch{\" o}ier}, F.~L., \& {van Dishoeck}, E.~F. 2005,
  \aap, 437, 501

\bibitem[{{Keane} {et~al.}(2001){Keane}, {Boonman}, {Tielens}, \& {van
  Dishoeck}}]{keane:so2}
{Keane}, J.~V., {Boonman}, A.~M.~S., {Tielens}, A.~G.~G.~M., \& {van Dishoeck},
  E.~F. 2001, \aap, 376, L5

\bibitem[{{Lada} {et~al.}(1984){Lada}, {Thronson}, {Smith}, {Schwartz}, \&
  {Glaccum}}]{lada:gl2591}
{Lada}, C.~J., {Thronson}, H.~A., {Smith}, H.~A., {Schwartz}, P.~R., \&
  {Glaccum}, W. 1984, \apj, 286, 302

\bibitem[{{Lahuis} \& {van Dishoeck}(2000)}]{lahuis:hcn}
{Lahuis}, F. \& {van Dishoeck}, E.~F. 2000, \aap, 355, 699

\bibitem[{{Linsky}(1998)}]{linsky:d}
{Linsky}, J.~L. 1998, Space Science Reviews, 84, 285

\bibitem[{{Lugo} {et~al.}(2004){Lugo}, {Lizano}, \& {Garay}}]{lugo:photevap}
{Lugo}, J., {Lizano}, S., \& {Garay}, G. 2004, \apj, 614, 807

\bibitem[{{Maret} {et~al.}(2002){Maret}, {Ceccarelli}, {Caux}, {Tielens}, \&
  {Castets}}]{maret:h2o}
{Maret}, S., {Ceccarelli}, C., {Caux}, E., {Tielens}, A.~G.~G.~M., \&
  {Castets}, A. 2002, \aap, 395, 573

\bibitem[{{Maret} {et~al.}(2005){Maret}, {Ceccarelli}, {Tielens}, {Caux},
  {Lefloch}, {Faure}, {Castets}, \& {Flower}}]{maret:ch3oh}
{Maret}, S., {Ceccarelli}, C., {Tielens}, A.~G.~G.~M., {et~al.} 2005, \aap,
  442, 527

\bibitem[{{Mitchell} {et~al.}(1995){Mitchell}, {Lee}, {Maillard}, {Matthews},
  {Hasegawa}, \& {Harris}}]{mitchell:gl490}
{Mitchell}, G.~F., {Lee}, S.~W., {Maillard}, J., {et~al.} 1995, \apj, 438, 794

\bibitem[{{Mitchell} {et~al.}(1990){Mitchell}, {Maillard}, {Allen}, {Beer}, \&
  {Belcourt}}]{mitchell:hot+cold}
{Mitchell}, G.~F., {Maillard}, J., {Allen}, M., {Beer}, R., \& {Belcourt}, K.
  1990, \apj, 363, 554

\bibitem[{{Mitchell} {et~al.}(1991){Mitchell}, {Maillard}, \&
  {Hasegawa}}]{mitchell:episodic}
{Mitchell}, G.~F., {Maillard}, J.-P., \& {Hasegawa}, T.~I. 1991, \apj, 371, 342

\bibitem[{{Ossenkopf} \& {Henning}(1994)}]{ossenkopf:opacities}
{Ossenkopf}, V. \& {Henning}, T. 1994, \aap, 291, 943

\bibitem[{{Pagani} {et~al.}(1992){Pagani}, {Wannier}, {Frerking}, {Kuiper},
  {Gulkis}, {Zimmermann}, {Encrenaz}, {Whiteoak}, {Destombes}, \&
  {Pickett}}]{pagani:h2d+}
{Pagani}, L., {Wannier}, P.~G., {Frerking}, M.~A., {et~al.} 1992, \aap, 258,
  472

\bibitem[{{Parise} {et~al.}(2005){Parise}, {Caux}, {Castets}, {Ceccarelli},
  {Loinard}, {Tielens}, {Bacmann}, {Cazaux}, {Comito}, {Helmich}, {Kahane},
  {Schilke}, {van Dishoeck}, {Wakelam}, \& {Walters}}]{parise:hdo}
{Parise}, B., {Caux}, E., {Castets}, A., {et~al.} 2005, \aap, 431, 547

\bibitem[{{Parise} {et~al.}(2003){Parise}, {Simon}, {Caux}, {Dartois},
  {Ceccarelli}, {Rayner}, \& {Tielens}}]{parise:solid}
{Parise}, B., {Simon}, T., {Caux}, E., {et~al.} 2003, \aap, 410, 897

\bibitem[{{Phillips} {et~al.}(1978){Phillips}, {Kwan}, {Scoville}, {Huggins},
  \& {Wannier}}]{phillips:h2o18}
{Phillips}, T.~G., {Kwan}, J., {Scoville}, N.~Z., {Huggins}, P.~J., \&
  {Wannier}, P.~G. 1978, \apjl, 222, L59

\bibitem[{{Phillips} {et~al.}(1992){Phillips}, {van Dishoeck}, \&
  {Keene}}]{phillips:h3o+}
{Phillips}, T.~G., {van Dishoeck}, E.~F., \& {Keene}, J. 1992, \apj, 399, 533

\bibitem[{{Poetzel} {et~al.}(1992){Poetzel}, {Mundt}, \& {Ray}}]{poetzel:h-h}
{Poetzel}, R., {Mundt}, R., \& {Ray}, T.~P. 1992, \aap, 262, 229

\bibitem[{{Pontoppidan} {et~al.}(2003){Pontoppidan}, {Fraser}, {Dartois},
  {Thi}, {van Dishoeck}, {Boogert}, {d'Hendecourt}, {Tielens}, \&
  {Bisschop}}]{pontoppidan:co_ice}
{Pontoppidan}, K.~M., {Fraser}, H.~J., {Dartois}, E., {et~al.} 2003, \aap, 408,
  981

\bibitem[{{Pontoppidan} {et~al.}(2004){Pontoppidan}, {van Dishoeck}, \&
  {Dartois}}]{pontoppidan:ice_map}
{Pontoppidan}, K.~M., {van Dishoeck}, E.~F., \& {Dartois}, E. 2004, \aap, 426,
  925

\bibitem[{{Preibisch} {et~al.}(2003){Preibisch}, {Balega}, {Schertl}, \&
  {Weigelt}}]{preibisch:afgl2591}
{Preibisch}, T., {Balega}, Y.~Y., {Schertl}, D., \& {Weigelt}, G. 2003, \aap,
  412, 735

\bibitem[{{Schilke} {et~al.}(1997){Schilke}, {Groesbeck}, {Blake}, \&
  {Phillips}}]{schilke:345survey}
{Schilke}, P., {Groesbeck}, T.~D., {Blake}, G.~A., \& {Phillips}, T.~G. 1997,
  \apjs, 108, 301

\bibitem[{{Sch\"{o}ier} {et~al.}(2005){Sch\"{o}ier}, {van der Tak}, {van
  Dishoeck}, \& Black}]{schoeier:moldata}
{Sch\"{o}ier}, F.~L., {van der Tak}, F.~F.~S., {van Dishoeck}, E.~F., \& Black,
  J.~H. 2005, \aap, 432, 369

\bibitem[{{Schreyer} {et~al.}(2002){Schreyer}, {Henning}, {van der Tak},
  {Boonman}, \& {van Dishoeck}}]{schreyer:gl490}
{Schreyer}, K., {Henning}, T., {van der Tak}, F.~F.~S., {Boonman}, A.~M.~S., \&
  {van Dishoeck}, E.~F. 2002, \aap, 394, 561

\bibitem[{{Snell} {et~al.}(2000){Snell}, {Howe}, {Ashby}, {Bergin}, {Chin},
  {Erickson}, {Goldsmith}, {Harwit}, {Kleiner}, {Koch}, {Neufeld}, {Patten},
  {Plume}, {Schieder}, {Stauffer}, {Tolls}, {Wang}, {Winnewisser}, {Zhang}, \&
  {Melnick}}]{snell:swas}
{Snell}, R.~L., {Howe}, J.~E., {Ashby}, M.~L.~N., {et~al.} 2000, \apjl, 539,
  L101

\bibitem[{{Stark} {et~al.}(1999){Stark}, {van der Tak}, \& {van
  Dishoeck}}]{stark:h2d+}
{Stark}, R., {van der Tak}, F.~F.~S., \& {van Dishoeck}, E.~F. 1999, \apjl,
  521, L67

\bibitem[{{Tamura} \& {Yamashita}(1992)}]{tamura:gl2591}
{Tamura}, M. \& {Yamashita}, T. 1992, \apj, 391, 710

\bibitem[{{Trinidad} {et~al.}(2003){Trinidad}, {Curiel}, {Cant{\' o}},
  {D'Alessio}, {Rodr{\'{\i}}guez}, {Torrelles}, {G{\' o}mez}, {Patel}, \&
  {Ho}}]{trinidad:gl2591}
{Trinidad}, M.~A., {Curiel}, S., {Cant{\' o}}, J., {et~al.} 2003, \apj, 589,
  386

\bibitem[{Van~der Tak(2005)}]{vdtak:catania}
Van~der Tak, F.~F.~S. 2005, in IAU Symposium 227 -- Massive Star Birth: A
  Crossroads of Astrophysics, ed. R.~Cesaroni, 70--78

\bibitem[{{Van der Tak} {et~al.}(2003){Van der Tak}, {Boonman}, {Braakman}, \&
  {van Dishoeck}}]{vdtak:sulphur}
{Van der Tak}, F.~F.~S., {Boonman}, A.~M.~S., {Braakman}, R., \& {van
  Dishoeck}, E.~F. 2003, \aap, 412, 133

\bibitem[{{Van der Tak} \& {Menten}(2005)}]{vdtak:qband}
{Van der Tak}, F.~F.~S. \& {Menten}, K.~M. 2005, \aap, 437, 947

\bibitem[{{Van der Tak} {et~al.}(2005){Van der Tak}, {Neufeld}, {Yates},
  {Hogerheijde}, {Bergin}, {Sch{\" o}ier}, \& {Doty}}]{vdtak:h2o-benchmark}
{Van der Tak}, F.~F.~S., {Neufeld}, D., {Yates}, J., {et~al.} 2005, in The
  Dusty and Molecular Universe: A Prelude to Herschel and ALMA, ed. A.~Wilson,
  431--432

\bibitem[{{Van der Tak} \& {van Dishoeck}(2000)}]{vdtak:zeta}
{Van der Tak}, F.~F.~S. \& {van Dishoeck}, E.~F. 2000, \aap, 358, L79

\bibitem[{{Van der Tak} {et~al.}(2000{\natexlab{a}}){Van der Tak}, {van
  Dishoeck}, \& {Caselli}}]{vdtak:meth}
{Van der Tak}, F.~F.~S., {van Dishoeck}, E.~F., \& {Caselli}, P.
  2000{\natexlab{a}}, \aap, 361, 327

\bibitem[{{Van der Tak} {et~al.}(1999){Van der Tak}, {van Dishoeck}, {Evans},
  {Bakker}, \& {Blake}}]{vdtak:gl2591}
{Van der Tak}, F.~F.~S., {van Dishoeck}, E.~F., {Evans}, N.~J., {Bakker},
  E.~J., \& {Blake}, G.~A. 1999, \apj, 522, 991

\bibitem[{{Van der Tak} {et~al.}(2000{\natexlab{b}}){Van der Tak}, {van
  Dishoeck}, {Evans}, \& {Blake}}]{vdtak:massive}
{Van der Tak}, F.~F.~S., {van Dishoeck}, E.~F., {Evans}, N.~J., \& {Blake},
  G.~A. 2000{\natexlab{b}}, \apj, 537, 283

\bibitem[{{Van Dishoeck}(1998)}]{vdishoeck:faraday}
{Van Dishoeck}, E.~F. 1998, Faraday Discussions, 109, 31

\bibitem[{{Walmsley} \& {van der Tak}(2005)}]{walmsley:paris}
{Walmsley}, C.~M. \& {van der Tak}, F. F.~S. 2005, in The Dusty and Molecular
  Universe: A Prelude to Herschel and ALMA, ed. A.~Wilson, 55--60

\bibitem[{{Whitney} {et~al.}(2003){Whitney}, {Wood}, {Bjorkman}, \&
  {Wolff}}]{whitney:geometry}
{Whitney}, B.~A., {Wood}, K., {Bjorkman}, J.~E., \& {Wolff}, M.~J. 2003, \apj,
  591, 1049

\bibitem[{{Wyrowski} {et~al.}(1999){Wyrowski}, {Schilke}, {Walmsley}, \&
  {Menten}}]{wyrowski:w3(h2o)}
{Wyrowski}, F., {Schilke}, P., {Walmsley}, C.~M., \& {Menten}, K.~M. 1999,
  \apjl, 514, L43

\end{thebibliography}

\end{document}